\documentclass[12pt,preprint]{aastex}
\usepackage{epsf}

\def\linebreak{\hfil\break}

\def\
{\hfil\linebreak}

\def\orch{{\it Orchestra}}
\def\deg{\ifmmode {^\circ}\else {$^\circ$}\fi}
\def\degree{\ifmmode {^\circ}\else {$^\circ$}\fi}
\def\mum{\ifmmode {\rm \mu {\rm m}}\else $\rm \mu {\rm m}$\fi}
\def\arcsec{\ifmmode ^{\prime \prime}\else $^{\prime \prime}$\fi}

\def\inch{\ifmmode ^{\prime \prime}\else $^{\prime \prime}$\fi}
\def\arcmin{\ifmmode ^{\prime}\else $^{\prime}$\fi}
\def\qprime{\ifmmode q^{\prime}\else $q^{\prime}$\fi}

\def\degree{\ifmmode {^\circ}\else {$^\circ$}\fi}
\def\mum{\ifmmode {\rm \mu {\rm m}}\else $\rm \mu {\rm m}$\fi}
\def\arcsec{\ifmmode ^{\prime \prime}\else $^{\prime \prime}$\fi}

\def\inch{\ifmmode ^{\prime \prime}\else $^{\prime \prime}$\fi}
\def\arcmin{\ifmmode ^{\prime}\else $^{\prime}$\fi}

\def\mjup{\ifmmode { M_J}\else $ M_J$\fi}
\def\rjup{\ifmmode { R_J}\else $ R_J$\fi}
\def\mearth{\ifmmode { M_{\oplus}}\else $ M_{\oplus}$\fi}
\def\rearth{\ifmmode { R_{\oplus}}\else $ R_{\oplus}$\fi}
\def\lstar{\ifmmode { L_d / L_{\star}}\else $ L_d / L_{\star}$\fi}
\def\ldlstar{\ifmmode { L_{\star}}\else $ L_{\star}$\fi}
\def\lsun{\ifmmode { L_{\odot}}\else $ L_{\odot}$\fi}
\def\mstar{\ifmmode M_{\star}\else $ M_{\star}$\fi}
\def\msun{\ifmmode M_{\odot}\else $ M_{\odot}$\fi}
\def\tstar{\ifmmode T_{\star}\else $ T_{\star}$\fi}
\def\rstar{\ifmmode R_{\star}\else $ R_{\star}$\fi}
\def\rsun{\ifmmode R_{\odot}\else $ R_{\odot}$\fi}
\def\mjup{\ifmmode M_{J}\else $ M_{J}$\fi}
\def\rjup{\ifmmode R_{J}\else $ R_{J}$\fi}
\def\mjupyr{\ifmmode { M_J~yr^{-1}}\else $ M_J~yr^{-1}$\fi}
\def\msunyr{\ifmmode { M_{\odot}~yr^{-1}}\else $ M_{\odot}~yr^{-1}$\fi}
\def\gyr{\ifmmode {\rm g~yr^{-1}}\else $\rm g~yr^{-1}$\fi}
\def\kms{\ifmmode {\rm km~s^{-1}}\else $\rm km~s^{-1}$\fi}
\def\ms{\ifmmode {\rm m~s^{-1}}\else $\rm m~s^{-1}$\fi}
\def\rhill{\ifmmode R_H\else $R_H$\fi}
\def\rfast{\ifmmode R_{fast}\else $R_{fast}$\fi}
\def\rgap{\ifmmode R_{gap}\else $R_{gap}$\fi}
\def\vhill{\ifmmode v_H\else $v_H$\fi}
\def\qdstar{\ifmmode Q_D^\star\else $Q_D^\star$\fi}
\def\rmax{\ifmmode r_{max}\else $r_{max}$\fi}
\def\r0{\ifmmode r_{0}\else $r_{0}$\fi}
\def\xm{\ifmmode x_{m}\else $x_{m}$\fi}
\def\gyr{\ifmmode {\rm g~yr^{-1}}\else ${\rm g~yr^{-1}}$\fi}
\def\cms{\ifmmode {\rm cm~s^{-1}}\else ${\rm cm~s^{-1}}$\fi}
\def\gcms{\ifmmode {\rm g~cm^{-2}}\else $\rm g~cm^{-2}$\fi}
\def\gcmc{\ifmmode {\rm g~cm^{-3}}\else $\rm g~cm^{-3}$\fi}

\def\2470{[24]--[70]}

\newbox\grsign \setbox\grsign=\hbox{$>$} \newdimen\grdimen \grdimen=\ht\grsign
\newbox\simlessbox \newbox\simgreatbox
\setbox\simgreatbox=\hbox{\raise.5ex\hbox{$>$}\llap
     {\lower.5ex\hbox{$\sim$}}}\ht1=\grdimen\dp1=0pt
\setbox\simlessbox=\hbox{\raise.5ex\hbox{$<$}\llap
     {\lower.5ex\hbox{$\sim$}}}\ht2=\grdimen\dp2=0pt

\def\simless{\mathrel{\copy\simlessbox}}

\begin{document}

\title{Coagulation Calculations of Icy Planet Formation Around 0.1--0.5~\msun\ Stars:
Super-Earths From Large Planetestimals
}
\vskip 7ex
\author{Scott J. Kenyon}
\affil{Smithsonian Astrophysical Observatory,
60 Garden Street, Cambridge, MA 02138} 
\email{e-mail: skenyon@cfa.harvard.edu}

\author{Benjamin C. Bromley}
\affil{Department of Physics, University of Utah, 
201 JFB, Salt Lake City, UT 84112} 
\email{e-mail: bromley@physics.utah.edu}
%
%-------------------------- ABSTRACT ----------------------------------
%
%\doublespace

\begin{abstract}

We investigate formation mechanisms for icy super-Earth mass planets orbiting
at 2--20~AU around 0.1--0.5 \msun\ stars. A large ensemble of coagulation 
calculations demonstrates a new formation channel: disks composed of large 
planetesimals with radii of 30--300~km form super-Earths on time scales 
of $\sim$ 1~Gyr.  In other gas-poor disks, a collisional cascade grinds 
planetesimals to dust before the largest planets reach super-Earth masses. 
Once icy Earth-mass planets form, they migrate through the leftover swarm of 
planetesimals at rates of 0.01--1~AU~Myr$^{-1}$. On time scales of 10~Myr 
to 1~Gyr, many of these planets migrate through the disk of leftover 
planetesimals from semimajor axes of 5--10 AU to 1--2 AU. A few per cent of 
super-Earths might migrate to semimajor axes of 0.1--0.2~AU.  When the disk
has an initial mass comparable with the minimum mass solar nebula scaled to 
the mass of the central star, the predicted frequency of super-Earths matches 
the observed frequency.

\end{abstract}

\keywords{Planetary systems -- Planets and satellites: formation -- 
Planets and satellites: physical evolution -- planet disk interactions -- 
stars: low mass}

\section{INTRODUCTION}
\label{sec: intro}

Super-Earths are among the most common planets known in the galaxy. 
With typical masses $m_p \approx$ 1--10 \mearth\ \citep[e.g.,][]{haghi2011,haghi2013}, 
super-Earths have been identified far from (microlensing, direct imaging) and 
close to (radial velocities, transits) their host stars 
\citep[e.g.,][]{rivera2005,beau2006,forv2009,gould2010,currie2012,bonfils2013}.
At any orbital separation, super-Earths are much more common than ice giants 
($m_p \approx$ 10--50 \mearth) or gas giants 
($m_p \approx$ 50--6000 \mearth), but they are probably less common than 
Mars-mass or Earth-mass planets 
\citep[e.g.,][]{gould2010,youdin2011,dress2013,fress2013,koppa2013,petit2013}.

Super-Earths orbiting stars with masses \mstar\ $\approx$ 0.1--0.5 \msun\ are 
an interesting challenge for planet formation theory.  At birth, most -- if not 
all -- low mass stars are surrounded by large circumstellar disks of gas and 
dust \citep[e.g.,][]{kh1995,scholz2006,luhman2012}. Disk masses derived from mm 
observations are typically 0.1\% to 10\% of the stellar mass 
\citep[e.g.,][]{will2011,and2013}. 
In a solar metallicity disk surrounding a 0.1~\msun\ star, the maximum mass in solid 
material is roughly 30~\mearth. Thus, super-Earths orbiting low mass stars 
contain a large fraction of the solid material originally in the disk.

Here, we consider super-Earth formation in the context of the planetesimal theory.
In this picture, dust grains in the circumstellar disk settle out of the gas and
grow into roughly km-sized or larger planetesimals \citep[e.g.,][]{chiang2010}.
Mergers among colliding planetesimals produce larger and larger objects, concentrating
solid material into a smaller and smaller number of massive protoplanets
\citep[e.g.,][]{saf1969,weth1989,gold2004}. Eventually, collisions and gravitational 
interactions among a few remaining protoplanets produce a set of planets on stable 
orbits \citep[e.g.,][]{chambers1998,raymond2004,kb2006,raymond2011}.

To maximize the probability of super-Earth formation, we examine numerical coagulation 
calculations of the growth of icy planetesimals at large distances from the parent star. 
We consider systems where the $e$-folding time for the gas surface density is 1~Myr.
In massive disks composed of very large planetesimals with $r \gtrsim$ 30~km, we
identify a new formation path: super-Earths from collisions of sub-Earth-mass objects 
at late times, $\sim$ 1--10~Gyr. This time scale is much longer than typical times,
$\sim$ 1--10 Myr, inferred from previous studies of formation within longer-lived gaseous 
disks \citep[e.g.,][]{laugh2004,koba2010b,koba2011,rogers2011}.  Because the accretion 
process and formation times for these two paths are so different, observations can 
test whether either path leads to super-Earths.

We describe the physics included in the coagulation calculations in \S2,
the outcomes of the calculations in \S3, and the ability of super-Earths
to migrate through the disk in \S4. In \S5, we compare our results to 
previous investigations and evaluate observational tests of model 
predictions. We conclude in \S6 with a brief summary.

\section{PLANET FORMATION CALCULATIONS}
\label{sec: calcs}

To calculate the formation and evolution of planets around low mass stars, we use \orch, 
an ensemble of computer codes for the formation and evolution of planetary systems. 
Currently, \orch\ consists of a radial diffusion code which computes the time evolution 
of a gaseous or a particulate disk \citep{bk2011a}, a multiannulus coagulation code which 
computes the time evolution of a swarm of planetesimals \citep{kb2004a,kb2008,kb2012}, 
and an $n$-body code which computes the orbits of gravitationally interacting protoplanets 
\citep{bk2006,bk2011b,bk2013}.  Within the coagulation code, \orch\ includes algorithms for 
treating interactions between small particles and the gaseous disk 
\citep[e.g.,][]{ada76,weiden1977a} and between coagulation mass bins and $n$-bodies
\citep{bk2006}. To treat interactions between small particles and the $n$-bodies more
rigorously, the $n$-body calculations include tracer particles. Massless tracer particles
allow us to calculate the evolution of the orbits of small particles in response to the
motions of massive protoplanets. Massless and massive tracer particles enable calculations 
of the response of $n$-bodies to the changing gravitational potential of small particles
\citep{bk2011a, bk2011b, bk2013}.

We perform coagulation calculations on a cylindrical grid with inner radius 
$a_{in}$ and outer radius $a_{out}$ surrounding a star with \mstar\ = 0.1, 
0.3, or 0.5 \msun. The model grid contains $N$ concentric annuli with widths 
$\delta a_i = 0.025 a_i$ centered at semimajor axes $a_i$.  To connect this 
study with previous analyses of dust production at $a$ = 30--150~AU around 
1--3~\msun\ stars \citep[e.g.,][]{kb2008,kb2010}, we set the equilibrium 
blackbody temperature of a dust grain at the inner edge of the grid at 
60--70~K for stellar ages of roughly 10~Myr \citep{bac1993,bar1998}.  Thus,
the inner edge of the grid is at $a_{in}$ = 7~AU (0.5~\msun), 4~AU (0.3~\msun), 
and 2.5~AU (0.1~\msun).  Although not ideal for predicting the frequency 
of debris disks at short wavelengths, grains with temperatures of 30--60~K 
are detectable with {\it Spitzer} \citep[e.g.,][]{rieke2005,trill2008} and 
{\it Herschel} \citep[e.g.,][]{matt2010,lest2012,eiroa2013}. We plan to discuss 
predictions for debris disks around 0.1--0.5~\msun\ stars in a separate paper.

To complete calculations in a reasonable amount of cpu time, we set $N$ = 64 
which yields $a_{out}/a_{in} \approx$ 5.  With $a_{out}/a_{in}$ fixed, choosing 
larger $a_{in}$ allows disks with larger masses. However, results for 
1--3~\msun\ stars from \citet{kb2008,kb2010} suggest very long formation times
at $a \gtrsim$ 40~AU (15~AU) for 0.5~\msun\ (0.1~\msun) stars.  Adopting 
$a_{in} \lesssim$ 3~AU (1~AU) for 0.5~\msun\ (0.1~\msun) stars yields formation 
times of $\lesssim$ 1~Myr, enabling better tests of theories for gas giant planet 
formation \citep[e.g.,][]{kk2008,koba2010b,koba2011}. However, smaller disks have 
smaller total masses. With our focus on super-Earth formation, a larger disk allows 
us to explore new regions of parameter space compared to previous calculations.

Within every annulus $i$, there are $M$ mass batches with characteristic mass 
$m_{ik}$ and radius $r_{ik}$ \citep{weth1993,kl1998}. Batches are logarithmically
spaced in mass, with mass ratio $\delta \equiv m_{ik+1} / m_{ik}$. In these
calculations $\delta = 2.0$ for $r_{ik} \lesssim$ 1~km; $\delta = 1.4$ for 
$r_{ik} \gtrsim$ 1~km.  For a handful of mass batches with $r_{ik} \approx$ 
1~km, $\delta$ makes a smooth transition from 2.0 to 1.4.  Each mass batch 
contains $N_{ik}$ particles with total mass $M_{ik}$. Thus, the average mass 
in a batch is $\langle m_{ik} \rangle = M_{ik} / N_{ik}$. Throughout the calculation,
this average mass is used to calculate the collision cross-section, collision 
energy, and other necessary physical variables. As mass is added and removed
from each batch, the average mass changes \citep{weth1993}.

For any $\delta$, numerical calculations lag the result of an ideal calculation
with infinite mass resolution.  We quantify this delay with (i) a suite of 
calculations using a cross-section with an analytic solution and (ii) a small
set of calculations with our standard cross-sections \citep[e.g.,][]{weth1990}.
When the cross-section is the product of planetesimal masses, the largest
protoplanet becomes a runaway object when it begins to grow by accreting the
next largest protoplanet \citep{weth1990}. This runaway begins at a 
dimensionless time $\eta$ = 1.0. Calculations with $\delta$ = 1.25 yield a 
runaway object at $\eta = 1.012$, lagging the ideal calculation by 1.2\% 
\citep[see also][]{kl1998}. Other simulations have lags of 0.5\% 
($\delta = 1.08$), 2.7\% ($\delta = 1.4$), and 8.7\% ($\delta = 2.0$).

To derive lags for the cross-sections used in our main calculations (see below), 
we consider simulations of 1~m to 1~km planetesimals in a single annulus at 7~AU 
around a 0.5~\msun\ star. These simulations do not include fragmentation.  
To establish a baseline result for $\delta$ = 1.08, we derive the average time 
$t_{1.08}$ and its dispersion $\sigma_t$ for the formation of the first object 
with a mass of $10^{25}$~g in a suite of twenty simulations.  We repeat this 
exercise for larger $\delta$ and derive the typical lag $(t_\delta - t_{1.08})/t_{1.08}$.  
For simulations with constant $\delta$ for all mass bins, we infer lags of 
0.5\% ($\delta$ = 1.25), 1.9\% ($\delta$ = 1.4), and 5.1\% ($\delta$ = 2.0). 
The typical dispersion in the lag ranges from 0.9\% for $\delta$ = 1.08 to 
6.3\% for $\delta = 2.0$.  Because standard cross-sections have a weaker 
scaling with mass, these lags are slightly smaller than the analytic case.  
When $\delta$ varies smoothly from 2.0 to 1.4 as in our main calculations, 
the lag is similar to calculations with constant $\delta$ = 1.4, 2.0\% 
instead of 1.9\%. With cpu time scaling as $\delta^2$, adopting our variable 
$\delta$ saves cpu time with little loss of accuracy.

To estimate the impact of $\delta$ on the collisional cascade, we examine a
suite of simulations with fragmentation.  We consider two measures -- the 
slope $q^{\prime}$ of the cumulative size distribution from 1~m to 30~m and 
the total mass loss $M_f$ -- at the time when the first $10^{25}$~g object 
forms.  For both quantities, the dispersion in the average quantity grows 
with $\delta$.  The differences in the average quantities are comparable to 
the dispersion.  Thus, we see no obvious impact of $\delta$ on the collisional 
cascade.

For these suites of simulations, the gas damps the velocities of small particles
with radii $\sim$ 1--100~m \citep{weth1993,kl1999a}.  Although our ability to
resolve the transition from high velocities ($r_{ik} \gtrsim$ 10--30~M) to low 
velocities ($r_{ik} \lesssim$ 1--10~m) depends on $\delta$, the general behavior 
of the calculations is independent of $\delta$. When $10^{25}$~g protoplanets 
form on time scales of 1 Myr or shorter, the surface density of the gas is 
comparable to the initial surface density.  
As in \citet[][and references therein]{koba2012}, velocity damping by the gas 
reduces velocities of particles with sizes of 1--10~m to less than a few m~s$^{-1}$ 
($e \lesssim 10^{-3}$). At these low velocities, collisions produce mergers 
instead of disruptions. Halting the collisional cascade leads to an excess of 
particles at these sizes.  When $10^{25}$~g protoplanets form on time scales 
of 1--2 Myr or longer, the surface density of the gas declines as the collisional
cascade begins. In these cases, gas is less effective at damping the velocities 
of small particles and halting the collisional cascade.  Thus, there is little 
or no excess of particles at these sizes. 

The calculations described in \S3 begin with a cumulative size distribution 
$N_c(>r_{ik}) \propto r_{ik}^{-q^{\prime}}$ of planetesimals with mass density
$\rho_p$ = 1.5 g cm$^{-3}$ and maximum initial size $r_0$.  For
comparison with investigators that quote differential mass
distributions with $N(r) \propto r^{-q}$, $q^{\prime}$ = $q$ - 1.
Here, we adopt an initial $q^{\prime}$ = 0.5 (most of the mass in
large objects) or 3.0 (equal mass per logarithmic bin).

Planetesimals have horizontal and vertical velocities $h_{ik}(t)$ and $v_{ik}(t)$ 
relative to a circular orbit.  The horizontal velocity  is related to the orbital 
eccentricity, $e_{ik}^2(t)$ = 1.6 $(h_{ik}(t)/V_{K,i})^2$, where $V_{K,i}$ is the 
circular orbital velocity in annulus $i$.  The orbital inclination depends on the 
vertical velocity, $({\rm sin}~i_{ik}(t))^2$ = $2(v_{ik}(t)/V_{K,i})^2$.

The mass and velocity distributions of the planetesimals evolve in time due to
inelastic collisions, drag forces, and gravitational encounters.  As summarized 
in \citet{kb2004a,kb2008}, we solve a coupled set of coagulation equations which
treats the outcomes of mutual collisions between all particles with mass $m_j$ 
in annuli $a_i$.  We adopt the particle-in-a-box algorithm, where the physical 
collision rate is $n \sigma v f_g$, $n$ is the number density of objects, $\sigma$ 
is the geometric cross-section, $v$ is the relative velocity, and $f_g$ is the 
gravitational focusing factor \citep{weth1993,kl1998}.  Depending on physical 
conditions in the disk, we derive $f_g$ in the dispersion or the shear regime 
\citep{kl1998,kb2012}.  For a specific mass bin, the solutions include terms 
for (i) loss of mass from mergers with other objects and (ii) gain of mass from 
collisional debris and mergers of smaller objects.

Once protoplanets reach radii of $\sim$ 2000~km, they acquire gaseous atmospheres.
Gas drag within the atmosphere slows down nearby small particles, enhancing the
accretion rate \citep[e.g.,][]{podo1988,kary1993}. To derived this enhanced rate,
we follow \citet{inaba2003} and solve for the hydrostatic structure of an atmosphere
with an outer radius equal to the smaller of the Bondi radius and the Hill radius.
With this solution, we calculate the cross-section as a function of the planetesimal
mass \citep[see also][]{chambers2008,koba2011,bk2011a}.

Collision outcomes depend on the ratio $Q_c/Q_D^*$, where $Q_D^*$ is the collision 
energy needed to eject half the mass of a pair of colliding planetesimals to infinity
and $Q_c$ is the center of mass collision energy 
\citep[see also][]{weth1993,will1994,tanaka1996b,stcol1997a,kl1999a,obrien2003,koba2010a}. 
Consistent with N-body simulations of collision outcomes 
\citep[e.g.,][]{benz1999,lein2008,lein2009}, we set
\begin{equation}
Q_D^* = Q_b r^{\beta_b} + Q_g \rho_p r^{\beta_g}
\label{eq:Qd}
\end{equation}
where $Q_b r^{\beta_b}$ is the bulk component of the binding energy,
$Q_g \rho_g r^{\beta_g}$ is the gravity component of the binding energy,
and $r$ is the radius of a planetesimal.

To explore the sensitivity of our results to the fragmentation algorithm, we 
consider two sets of parameters $f_i$.  As in \citet[][2010]{kb2008}, we set 
$\rho_p$ = 1.5~g~cm$^{-3}$ and adopt parameters for `strong' ($f_s$) and `weak' 
($f_w$) planetesimals, where \qdstar($f_s$) $>$ \qdstar($f_w$) for
$r \gtrsim$ 10~m \citep[see Fig. 1 in][]{kb2012}.  Strong planetesimals have 
$f_s$ = $\{Q_b$ = $10^3$ erg g$^{-1}$, $\beta_b$ = 0, $Q_g$ = 
1.5 erg g$^{-2}$~cm$^{1.75}$, $\beta_g$ = 1.25$\}$.  Weak planetesimals have 
$f_w$ = $\{ Q_b$ = $2 \times 10^5$ erg g$^{-1}$ cm$^{0.4}$, $\beta_b = -0.40$, 
$Q_g$ = 0.22 erg g$^{-2}$ cm$^{1.7}$, $\beta_g$ = 1.30 $\}$.

These parameters are broadly consistent with published analytic and numerical
simulations \citep[e.g.,][]{davis1985,hols1994,love1996,housen1999}.  At small 
sizes, they agree with results from laboratory experiments of impacts with icy 
targets and projectiles \citep[see also][]{ryan1999,arakawa2002,giblin2004,burchell2005}. 
In the gravity regime, our parameters agree with simulations from \citet{benz1999} 
for strong objects and \citet{lein2009} for weak objects.  In the strength regime, 
our choice for weak planetesimals follows \citet{lein2009}; our choice for strong 
planetesimals allows us to test whether outcomes depend on $\beta_b$ 
\citep[see also][]{kb2008,kb2010}.

For two colliding 
planetesimals with masses $m_1$ and $m_2$, the mass of the merged planetesimal is
\begin{equation}
m = m_1 + m_2 - m_{ej} ~ ,
\label{eq:msum}
\end{equation}
where the mass of debris ejected in a collision is
\begin{equation}
m_{ej} = 0.5 ~ (m_1 + m_2) \left ( \frac{Q_c}{Q_D^*} \right)^{9/8} ~ .
\label{eq:mej}
\end{equation}
In recent calculations of \citet{koba2010a} and \citet{koba2010b,koba2011,koba2012}, 
the exponent in an 
equivalent expression for the amount of ejected mass is 1 instead of our adopted 
9/8. Compared to \citet{koba2010a}, our calculations produce somewhat more debris 
in high velocity collisions ($Q_c > \qdstar$) and somewhat less debris in low 
velocity collisions ($Q_c < \qdstar$). For the calculations in this paper, the two 
approaches lead to roughly a factor of two difference in the amount of mass lost 
to high and low velocity collisions.  Given the factor of 2--3 (10 or more) 
differences in plausible coefficients for $Q_g$ ($Q_b$), this uncertainty is 
within the margin of error for these types of calculations. 

To place the debris in our grid of mass bins, we set the mass of the largest collision
fragment as $m_L = 0.2 m_{esc}$ and adopt a cumulative mass distribution
$N_c \propto m^{-q_d}$ with $q_d$ = 0.833, roughly the value expected for a system 
in collisional equilibrium \citep{dohn1969,will1994,tanaka1996b,obrien2003,koba2010a}. 
This approach allows us to derive ejected masses for catastrophic collisions
with $Q_c \sim Q_D^*$ and for cratering collisions with $Q_c \ll Q_D^*$
\citep[see also][]{weth1993,will1994,tanaka1996b,stcol1997a,kl1999a,obrien2003,koba2010a}. 

When fragmentation begins, the important physical quantity in the calculations is the 
ratio of two rates: the rate leftover planetesimals collide and fragment into smaller 
objects and the rate protoplanets accrete the leftovers and their fragments. When 
planetesimals and their fragments collide more often with other planetesimals, the 
collisional cascade removes fragments from the system more rapidly than protoplanets 
accrete them. The growth of protoplanets stalls \citep{inabawet2003,kb2004a}. 

Although variations in the fragmentation parameters have a large impact on the mass
and visibility of the smallest particles \citep[e.g.,][]{bely2011}, plausible variations 
in the parameters adopted for the size distribution of fragments have little impact on 
these two collision rates.  The rate protoplanets accrete leftover small planetesimals 
and fragments depends primarily on the total mass in small objects.  
Throughout the evolution, most of the mass in the leftovers is in large objects with 
$R \gtrsim$~1~km. This mass is independent of $m_L$ and $q_d$. Thus protoplanet accretion 
rates do not depend on $m_L$ and $q_d$.  Fragment production depends on the collision
velocities and the relative collision rates between (i) pairs of leftover planetesimals 
and (ii) a leftover planetesimal and another much smaller object 
\citep[e.g.,][and references therein]{koba2010a,bely2011}.  With most of the mass in 
large objects and a collision rate $\dot{n} \propto \Sigma(R) \Omega R^2$, the production 
rate of fragments (i) is independent of $m_L$ and $q_d$ for collisions of pairs of
planetesimals and (ii) depends weakly on $m_L$ and $q_d$ for collisions between a
large object and a much smaller object \citep{koba2010a}.  Thus, the adopted size 
distribution for the fragments has little impact on the growth of protoplanets.

To compute the evolution of the velocity distribution, we include collisional damping 
from inelastic collisions, gravitational interactions, and gas drag.  For inelastic and 
elastic collisions, we follow the statistical, Fokker-Planck approaches of \citet{oht1992} 
and \citet{oht2002}, which treat pairwise interactions (e.g., dynamical friction and viscous 
stirring) between all objects in all annuli.  As in \citet{kb2001}, we add terms to treat 
the probability that objects in annulus $i_1$ interact with objects in annulus 
$i_2$ \citep{kb2004b,kb2008}. We also compute long-range stirring from distant oligarchs 
\citep{weiden1989}. 

To derive damping and radial transport from gas drag, we include approaches developed in
\citet{weiden1977a} and \citet{raf2004}. Following \citet{weiden1977a}, we solve the equations
of motion for particles in the Epstein, Stokes, intermediate, and quadratic regimes.  The 
resulting drift velocity as a function of particle size allows us to calculate the rate
particles are lost through each annulus in the grid. The surface density of particles 
well-coupled to the gas diminishes exponentially with the gas depletion time defined below.
We use the scale-height of the gas to set their relative velocities \citep{bk2011a}.  For 
poorly-coupled particles, the surface density declines with the drift rate. To set particle 
velocities, we adopt the results of \citet{ada76}, who derive stirring rates for $e$ and 
$i$ in each regime.

Solving the equations of motion for all particles in each time step is computationally
intensive.  For time step $\Delta t$ and gas depletion time $t_{gas}$ defined below, we 
solve the equations of motion on a time scale $t_{eq}$, with $\Delta t \ll t_{eq} \ll t_{gas}$.
In between these evaluations, we follow the analytic approach of \citet{raf2004} scaled 
to yield similar solutions to the equations of motion as in the \citet{weiden1977a} 
approach. Once the gas surface density declines to 0.01\% of its initial velue, we 
discontinue gas drag calculations. Tests using only one approach result in nearly identical 
long-term evolution of the system. 

The initial conditions for these calculations are appropriate for a disk with an age 
of $\lesssim$ 1--2~Myr \citep[e.g.,][and references therein]{currie2007c,will2011}.  
We consider systems of $N = 64$ annuli in disks where the initial surface density 
of solid material follows a power law in semimajor axis,
\begin{equation}
\Sigma_{d,i} = \Sigma_{d,0}(M_{\star}) ~ x_m ~ a_i^{-k} ~ , 
\label{eq:sigma-dust}
\end{equation}
where $a_i$ is the central radius of the annulus in~AU, $k$ = 1 or 3/2, and
$x_m$ is a scaling factor.  For a standard gas to dust ratio of 100:1, 
$\Sigma_{gas,0} = 100 ~ \Sigma_{d,0} (M_{\star})$.  
To explore a range of disk masses similar to the observed range among 
the youngest stars, we consider 
$\Sigma_{d,0}$ = 30~($M_\star / M_\odot$)~g~cm$^{-2}$ and $x_m$ = 0.01--3. 
Disks with $x_m \approx$ 0.1 have masses similar to the median disk masses observed around 
young stars in nearby dark clouds \citep{and2005,will2011,and2013}. Somewhat larger 
scale factors, $x_m \approx$ 1, correspond to models of the minimum mass solar nebula 
(MMSN) models of \citet{weiden1977b} and \citet{hayashi1981} for solar-type stars. 

To evolve the gas in time, we consider a simple nebular model for the gas density.
We adopt a scale height $H_{gas}(a) = H_{gas,0} (a/a_0)^{1.125}$ \citep{kh1987}
and assume the gas surface density declines exponentially with time
\begin{equation}
\Sigma_{gas}(a,t) = \Sigma_{gas,0} ~ x_m ~ a^{-k} ~ e^{-t/t_{gas}}
\label{eq:sigma-gas}
\end{equation}
where $\Sigma_{gas,0}$ and $x_m$ are scaling factors and $t_{gas}$ = 1~Myr is the 
gas depletion time.  Calculations of viscous protostellar disks with a viscosity 
parameter $\alpha \approx 10^{-2} - 10^{-3}$ yield typical depletion times of
0.5--5~Myr \citep[e.g.,][]{cha2009,bk2011a}.  Observations of the lifetimes of 
accretion disks in pre-main sequence stars suggest depletion times ranging from
$\sim$ 1~Myr to $\sim$ 10~Myr \citep{kh1995,currie2009,kk2009,mama2009,will2011}. 

This short depletion time has a direct impact on the final masses of protoplanets 
\citep[e.g.,][]{gold2004,koba2010b,koba2011}.  Gas drag damps the velocities of small 
objects and removes them from the grid \citep[e.g.,][]{ada76,weiden1977a}.  Damping 
small particle velocities speeds up runaway growth; removing small particles slows 
down runaway growth. Our short gas depletion time slows down runaway growth but 
allows more mass for planet formation. Thus, our calculations are more likely to 
produce super-Earth mass planets and less likely to produce gas giants than 
calculations with a longer gas depletion time \citep[e.g.,][]{koba2011,koba2012}.

Table~\ref{tab: massgrid} summarizes the model grids. For each set of $x_m$, we choose the 
extent of the disk; the initial radius of the largest planetesimal, $r_0$ = 1--1000~km; 
the initial eccentricity $e_0 = 10^{-4}$ and inclination $i_0 = 5 \times 10^{-5}$; 
the power law exponent of the initial size distribution, 
$q_{init}^\prime$ = $3$ (equal mass per log interval in mass) or $q_{init} = 0.5$ 
(most of the mass in the largest objects);
the fragmentation parameters, $f_s$ or $f_w$; and the evolution time $t_{max}$.  To 
understand the possible range of outcomes, we repeat calculations 10--15 times 
with different random number seeds for each combination of initial conditions. 

Our calculations follow the time evolution of the mass and velocity distributions of 
objects with a range of radii, $r_{ij} = r_{min}$ to $r_{ij} = r_{max}$.  The upper 
limit $r_{max}$ is always larger than the largest object in each annulus.  Erosive
collisions produce objects with $r_{ij}$ $< r_{min}$, which are `lost' to the model 
grid.  Until a calculation produces objects with $r_{max} \gtrsim$ 1000~km, lost 
objects comprise a small fraction, $\lesssim$ 1\%, of the total mass in the grid. 

Once large objects form, the fate of lost objects depends on $\Sigma_g$ the surface 
density of the gas.  When $\Sigma_g$ is small, small particles with $r_{ij}$ $< r_{min}$ 
are more likely to be ground down into smaller and smaller objects than to collide 
with larger objects in the grid \citep[see][2002b, 2004a]{kb2002a}.  Thus, lost 
objects have little impact on the growth of the largest objects.

When $\Sigma_g$ is large, small particles with $r_{ij}$ $< r_{min}$ are well-coupled 
to the gas \citep[e.g.,][]{raf2004}. Velocity damping by the gas then prevents erosive 
collisions of small particles. If protoplanets can accrete small fragments more rapidly
than gas drag removes them, protoplanets can reach masses of 1--10~\mearth\ on short
time scales \citep{raf2004,kb2009,ormel2010a,ormel2010b,bk2011a,lamb2012}. However, 
several recent calculations suggest gas drag effectively damps the random velocities 
of 1--10~m particles, halting the collisional cascade and producing an excess of 
particles at these sizes \citep[e.g.,][]{koba2010b}.  Gas drag removes particles 
with $r_{ij} \lesssim$ 1~m before larger objects can accrete them. Although gas drag 
cannot rapidly remove larger particles, these objects do not contain much mass. Thus, 
damping by gas drag limits the formation of super-Earth and gas giant planets from 
fragments produced in a collisional cascade \citep{koba2010b,koba2011,koba2012}.

To consider a broad set of initial collisions with a limited amount of computer time, 
we set $r_{min}$ = 100 cm in all of our calculations. For simulations with 
$r_0 \gtrsim$ 3~km, erosive collisions begin after the gaseous disk has lost 40\% to 
90\% of its initial mass. With gas drag weakened, lost objects are then ground down 
into smaller and smaller objects which are eventually ejected from the system. 
These calculations yield a reasonably accurate estimate for the final masses of 
protoplanets. With smaller planetesimals, $r_0 \lesssim$ 1~km, erosion begins with 
a significant gas disk. These simulations somewhat underestimate the likely masses of 
the largest objects \citep[e.g.,][]{koba2011}. 

To connect the outcomes of coagulation calculations with the results of analytic theory, 
we write the collision rate in terms of the surface density $\Sigma$ and the angular 
velocity $\Omega$ of disk material \citep[see also][]{liss1987,kb2008,ormel2010c,yk2012}. 
Adopting $H$ as the vertical scale height of disk material and $m$ as the mass of a particle, 
the number density of particles is roughly $n \propto \Sigma / m H$. With $v \approx H \Omega$ 
and $f_g \approx (v_{esc,l} / v_s)^2$, where $v_{esc,l}$ is the escape velocity of typical 
large objects and $v_s$ is the characteristic speed of smaller bodies, the collision rate 
is
\begin{equation}
{ dN \over dt } \approx \Sigma \Omega {\sigma \over m} \left ( { v_{esc,l} \over v_s } \right )^2 ~ .
\label{eq: coll}
\end{equation}
A single particle experiences $dN/dt$ collisions per time interval.  Defining the collision 
time, $t_c = (dN/dt)^{-1}$, 
\begin{equation}
t_c \approx (\Sigma \Omega)^{-1} { m \over \sigma } \left ( { v_s \over v_{esc,l} } \right )^2~ .
\label{eq: t-coll}
\end{equation}
With $\Sigma = \Sigma_0 x_m a^{-k}$ ($k$ = 1--2) and $\Omega = (G \mstar / a^3)^{1/2} = \Omega_0 a^{-3/2}$,
\begin{equation}
t_c \approx (\Sigma_0 x_m \Omega_0)^{-1} { m \over \sigma } \left ( { v_s \over v_{esc,l} } \right )^2 ~ a^{k+3/2} ~ .
\label{eq: t-coll-fin}
\end{equation}
The collision time grows with decreasing disk mass ($\Sigma_0$) and with increasing
distance from the central star. This time sets the growth timescales for the largest 
solid objects in the disk.

\section{EVOLUTION OF THE LARGEST OBJECTS}
\label{sec: evol}

\subsection{Growth with a Broad Range of Initial Planetesimal Sizes}

We begin with a discussion of planet formation in disks composed of small planetesimals
in a power law size distribution with \qprime\ = 3 over an initial size range of 1~m to 1~km.  
In this size range, viscous stirring from the largest objects is initially weaker than
dynamical friction \citep{stewart:2000,gold2004}.  Thus, the two important dynamical 
processes at the start of each calculation are gas drag on the small objects and dynamical 
friction between large and small objects \citep[][and references therein]{yk2012}. Gas drag 
reduces the random velocities of small objects \citep{ada76,weiden1977a}. Dynamical friction 
raises the random velocities of the smallest objects and damps the random velocities of the 
largest objects \citep[e.g.,][and references therein]{weth1989,weth1993}. As the velocities 
of the large objects drop relative to the velocities of the small objects, viscous stirring 
begins to dominate the velocity evolution of the small objects \citep{stewart:2000,oht2002}.

When \qprime\ = 3 and $r_0$ = 1~km, every calculation begins with a short phase of 
orderly growth \citep[see also][]{kl1998,kb2008,kb2010,kb2012}. Initially, all 
planetesimals have the same random velocities, which are comparable to the escape 
velocity (and much larger than the Hill velocity) of 1~km planetesimals.  Thus, 
gravitational focusing factors are small, $f_g \approx$ 1. Growth is orderly: 
lower mass objects grow more rapidly than more massive objects 
\citep[see][and references therein]{saf1969,yk2012}. 
As orderly growth proceeds, the random velocities of all planetesimals `relax' to 
a rough equilibrium where (i) viscous stirring and gas drag achieve an approximate 
balance for the random velocities of the smallest objects and (ii) dynamical friction 
maintains an approximate equipartition in random kinetic energy as a function of 
particle size.
For the initial gas densities in these calculations, the `equilibrium' random 
velocity $v_{eq,s}$ of the smallest particles is much smaller than the escape 
velocity $v_{esc,l}$ of the largest objects \citep{raf2004}. With 
$f_g \propto (v_{esc,l} / v_{eq,s})^2$, gravitational focusing factors increase 
considerably as the random velocities of all planetesimals relax 
\citep[see also][and references therein]{kl1998,gold2004,kb2012}. As $f_g$ increases,
the largest objects begin to grow much more rapidly than smaller objects. Runaway 
growth begins.

During the runaway, growth concentrates more and more of the mass in the largest objects.
With more mass in larger objects, viscous stirring becomes more and more important in
raising the velocities of the smallest objects. Eventually, viscous stirring overcomes 
damping by gas drag. The random velocities of the smallest planetesimals grow to 
$v_{esc,l}$. Gravitational focusing factors decrease.  Runaway growth ends.

Following the runaway, growth again becomes orderly. Throughout the radial grid, 
the velocities of the smallest planetesimals are a significant fraction of the 
escape velocity of the largest object in an annulus. Thus, gravitational focusing
factors are small and fairly independent of the masses of the largest objects.
Orderly growth scales inversely with mass; thus, smaller large objects grow 
somewhat faster than the largest large objects. In this period of oligarchic 
growth \citep{kok1998,kok2000,ormel2010c}, the largest object in each annulus 
evolves to a common size.

Throughout oligarchic growth, the random velocities of the small objects continue 
to grow. Eventually, the collision energy for a pair of small objects approaches 
\qdstar. Collisions then yield copious amounts of debris instead of a merger. 
Collisions among debris particles produce additional debris, leading to a 
collisional cascade where the small objects are ground to dust 
\citep[e.g.,][]{dohn1969,will1994,kb2002a,kb2002b,wyatt2002,dom2003}. Because 
destructive collisions among small objects are more likely than collisions
between a large object and a small object, the start of the collisional cascade
effectively ends the growth of the largest objects
\citep[see also][]{inabawet2003,kb2004a,koba2010b}.  During the cascade, these 
objects reach a characteristic maximum size, \rmax, which depends on the
initial disk mass and the fragmentation parameters \citep{kb2008, kb2010, kb2012}.
Tables~\ref{tab: rad1}--\ref{tab: rad2} list \rmax\ for the full suite of 
calculations.

\begin{figure}
\includegraphics[width=6.5in]{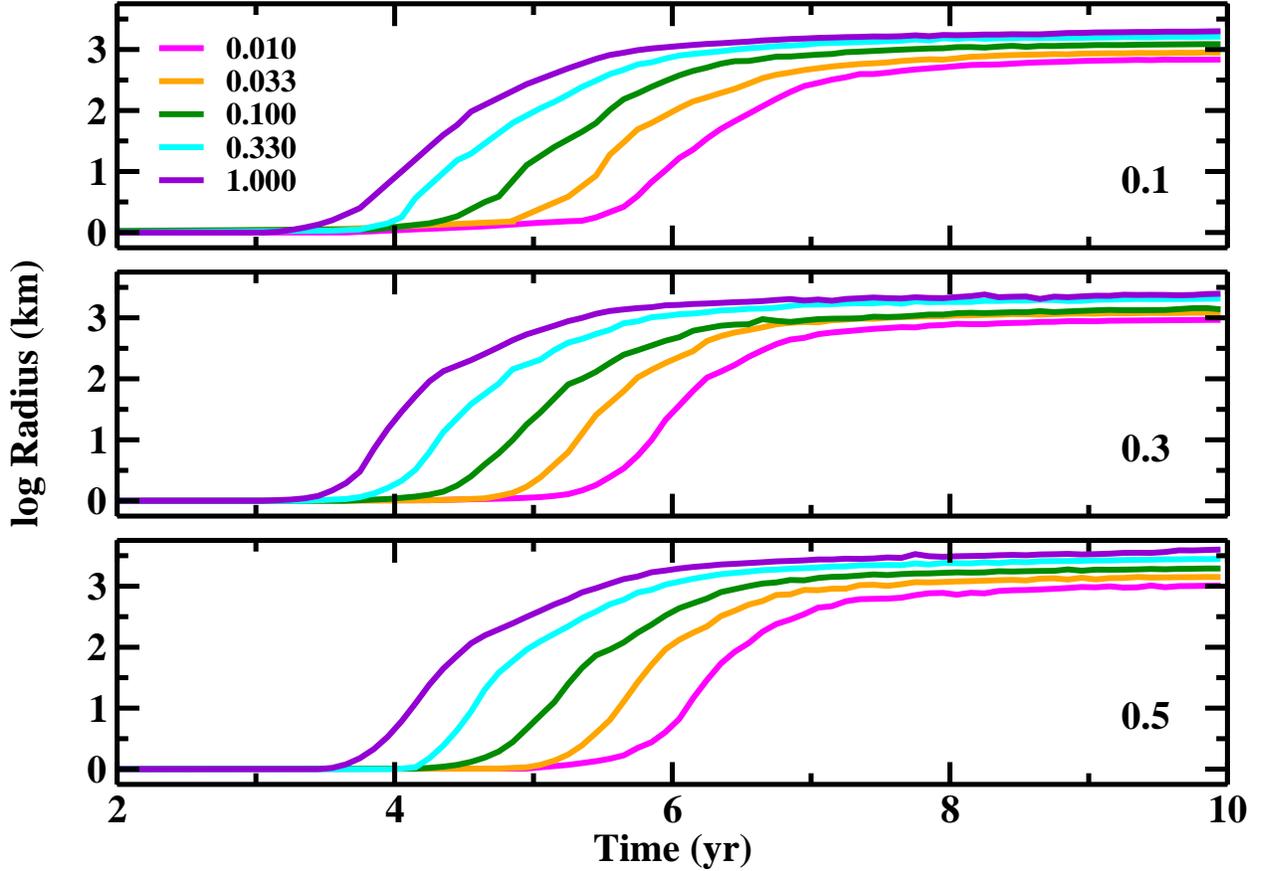}
\vskip 3ex
\caption{%
Growth of the largest object at 2.5~AU (upper panel, \mstar = 0.1 \msun), 
4~AU (middle panel, \mstar = 0.3 \msun), and 7~AU (lower panel, \mstar = 0.5 \msun)
for disks with $\Sigma \propto a^{-1}$, initial masses, $x_m$ = 0.01--1.0 (as 
indicated in the legend in the upper panel), and initial planetesimal radii 
$r_0$ = 1~km. During runaway growth, protoplanets grow rapidly from 1~km to 
roughly 300~km. As growth becomes oligarchic, a collisional cascade grinds leftover 
planetesimals to dust, robbing the protoplanets of material to accrete. Thus,
protoplanets slowly reach a typical maximum radius $r_{max} \approx$ 1000--3000~km.
\label{fig: rmax1}
}
\end{figure}

Fig.~\ref{fig: rmax1} illustrates this evolution for disks of various masses
around 0.1~\msun\ (upper panel), 0.3~\msun\ (middle panel), and 0.5~\msun\ (lower
panel) stars. The calculations begin with an initial surface density of solids
$\Sigma_d = 15 x_m (\mstar / 0.5 \msun) (a / {\rm 1~AU})^{-1}$~\gcms.  In each panel, 
the curves show the growth of the largest object within the innermost few annuli
of disks with the $f_s$ fragmentation parameters and \xm\ = 0.01--1 as indicated 
in the legend in the upper panel. As the velocity distribution relaxes at the 
start of each calculation, large objects begin their evolution with a short 
orderly growth phase where their radii grow by a few per cent.  When runaway 
growth begins, the largest objects grow in radius by factors of 100--300 in 
$10^4 - 10^6$~yr. As the radius approaches $\sim$ 300~km, gravitational 
focusing factors decline. Oligarchic growth begins.  During oligarchic growth, 
the collisional cascade removes more and more material from the grid. The 
sizes of the largest objects reach a constant value.

For all large objects, the growth time depends on the initial mass in the 
disk \citep[Fig.~\ref{fig: rmax1}, see also][]{kb2008,kb2010}. With 
$\Sigma \propto a^{-1}$ and $v_s / v_{esc,l} \approx 1$, the accretion 
time from eq. (\ref{eq: t-coll-fin}) is $t_c \propto x_m^{-1}$. When 
viscous stirring and gas drag are in an approximate equilibrium, 
$v_s / v_{esc,l} \propto \Sigma^{-\gamma_1}$, with $\gamma \approx$
1/6 to 1/5 (quadratic drag), 0 (Stokes drag) and 2/5 (Epstein drag)
\citep{raf2004}. \citet{koba2010b} derive similar results. This additional 
component to accretion implies $t_c \propto x_m^{-\gamma_2}$, 
with $\gamma_2 \approx$ 1.3--1.4 (quadratic drag), 1.0 (Stokes drag), and 
1.8 (Epstein drag) \citep[see also][Appendix]{kb2008}. 

With a typical $\gamma_2 \approx 1.1$ (see below), our results match 
predictions.  At the start of these calculations, the smallest (largest) 
particles are in the Epstein (quadratic) regime.  Intermediate-sized 
particles lie in the Stokes regime.  As the gas density drops with time, 
fewer and fewer particles are in the Epstein regime.  Gas drag, mergers 
among smaller objects, and accretion of small objects by large objects 
also remove mass from objects in the Epstein regime. Integrating over
the size distribution as the system evolves, 75\% to 85\% of the mass is 
in the Stokes or quadratic regime. Assuming $\gamma_2$ scales with mass
fraction, we expect $\gamma_2 \approx$ 1.05--1.2, close to our derived 
$\gamma_2 \approx$ 1.1. We plan to explore the relation between drag
regimes and $\gamma_2$ in a separate analysis.

\begin{figure}
\includegraphics[width=6.5in]{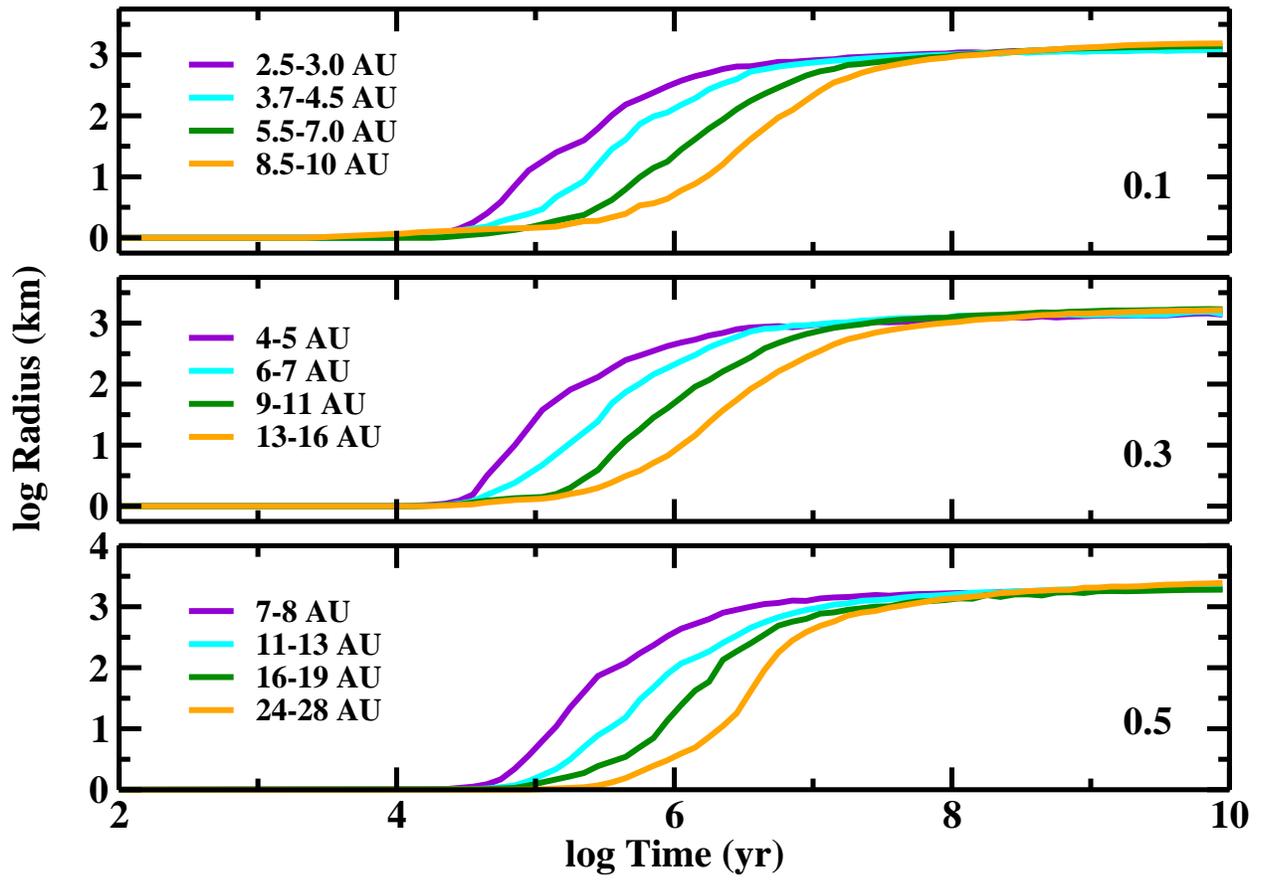}
\vskip 3ex
\caption{%
As in Fig. \ref{fig: rmax1} for different annuli in disks with $x_m$ = 0.1.
Protoplanets grow more rapidly in the inner disk.
\label{fig: rmax2}
}
\end{figure}

The growth rate also depends on the semimajor axis (Fig.~\ref{fig: rmax2}). In the 
inner disks of 0.1--0.5~\msun\ stars, protoplanets grow much more rapidly than at
much larger semimajor axes. For all initial disk masses, the time scale for the
radii of protoplanets to reach a fiducial radius scales as $t_c \propto a^{-5/2}$, 
the result predicted from analytic theory (eq.~[\ref{eq: t-coll}]).

Although the outcome of runaway growth is fairly insensitive to \qdstar, the 
final masses of protoplanets depend on the fragmentation parameters. In these
and other calculations with 1~m to 1~km planetesimals, runaway growth produces
a few large objects and leaves most of the initial mass in 0.1--10~km objects.
The collision energies $Q_c$ of these leftovers are comparable to the 
gravitational binding energies of the largest objects.  Because debris 
production scales with the ratio of $Q_c$ to \qdstar, weaker (stronger) 
leftovers produce substantial debris when the largest objects are smaller
(larger). Thus, the size of the largest objects (\rmax) correlates with 
\qdstar\ \citep[see also][]{inabawet2003,koba2010a,kb2010,kb2012,koba2011,koba2012}.

\begin{figure}
\includegraphics[width=6.5in]{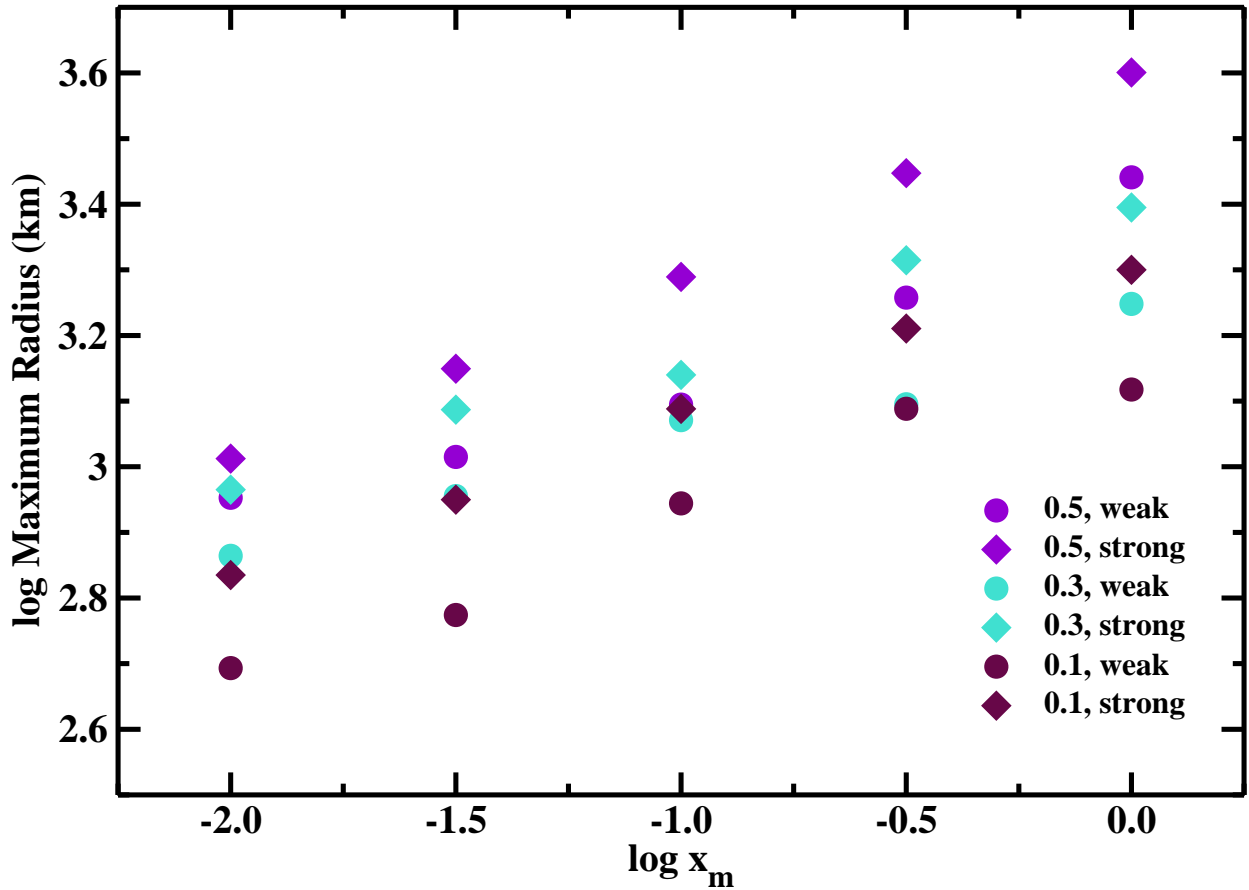}
\vskip 3ex
\caption{%
Evolution of maximum protoplanet radius with initial disk mass \xm\ for disks
with strong (filled diamonds) or weak (filled circles) planetesimals around
0.1 \msun\ (maroon points), 0.3 \msun\ (cyan points), or 0.5 \msun\ (violet points)
stars.  When 1~km planetesimals are stronger, protoplanets grow to larger sizes.
\label{fig: rmaxq}
}
\end{figure}

Fig.~\ref{fig: rmaxq} illustrates the variation of \rmax\ with the fragmentation
parameters. For identical initial conditions, the largest objects are 5\% to 20\% 
larger (20\% to 100\% more massive) when planetesimals are strong. Fragmentation
has a larger impact on the outcome in more massive disks. When $x_m \approx$ 1,
\rmax\ is 15\% to 20\% larger when planetesimals are strong. At $x_m \approx$ 0.01,
fragmentation has a 5\% to 10\% impact on \rmax.

\begin{figure}
\includegraphics[width=6.5in]{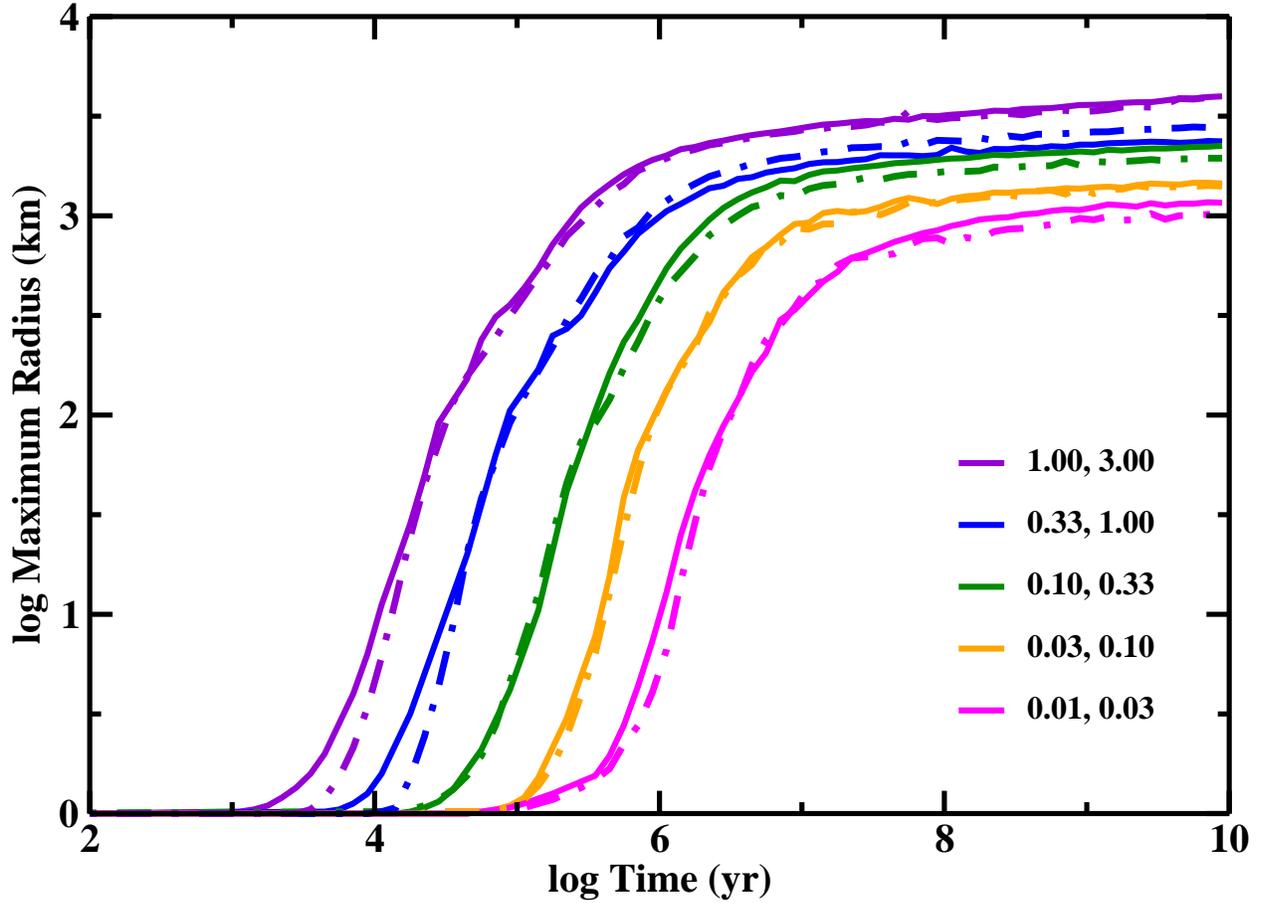}
\vskip 3ex
\caption{%
Evolution of maximum protoplanet radius in disks with $\Sigma \propto a^{-k}$ and
$k$ = 1 or 3/2 around a 0.5 \msun\ star. Solid curves indicate results for $k$ = 3/2;
dot-dashed curves show results for $k$ = 1. The legend indicates the initial disk mass
\xm\ for each power law index. The first (second) number in the legend indicates 
\xm\ for $k$ = 1 (3/2). 
\label{fig: rmaxn}
}
\end{figure}

Finally, protoplanet growth is fairly independent of the exponent in the initial 
power-law relation between the surface density of solids and semimajor axis
(Fig.~\ref{fig: rmaxn}). In calculations with similar total disk mass, the 
time scales for protoplanets to reach sizes of 300~km, 1000~km, and 3000~km
depend primarily on the mass in an annulus and the fragmentation parameters.
For fixed total mass, disks with $\Sigma \propto a^{-k}$ and $k$ = 3/2 have 
more material in the inner disk compared to disks with $k$ = 1. Thus, 
protoplanets grow faster in the inner (outer) part of the disk for 
$k$ = 3/2 (1). Over the entire disk, similar initial masses yield similar
\rmax.

To conclude this section on the growth of ensembles of 1~m to 1~km planetesimals
orbiting 0.1--0.5 \msun\ stars, we quote the median time scale for calculations
to produce objects with radii of 300~km ($t_{300}$) and 1000~km ($t_{1000}$) 
as a function of initial conditions. From eq. (\ref{eq: t-coll}), we expect
$t_c = t_0 x_m^{-\gamma2} a^{k + 3/2}$, where $t_0$ is a normalization factor
and $k$ is the exponent in the surface density law. For every calculation, we 
verify the $t_c \propto a^{k + 3/2}$ relation expected from theory. Our results 
match this expectation to within 1\% to 2\%. For every suite of 10--15 calculations 
with identical initial conditions, we derive the median time scale to reach
\rmax\ = 300~km or 1000~km. Among suites with identical fragmentation parameters,
we derive the best-fitting $t_0$ and $\gamma_2$ as a function of $a_{10}$,
where $a_{10} = a / {\rm 10~AU}$. This analysis yields $\gamma_2 = 1.1 \pm 0.02$ 
and the range of $t_0$ listed in Table~\ref{tab: times1}. For disks with masses 
similar to the MMSN and composed of 1~m to 1~km planetesimals, the time scale for 
protoplanets to grow to 300~km is roughly 0.1--1~Myr. Growth times for 1000~km 
protoplanets are roughly a factor of ten longer.

\subsection{Growth with a Narrow Range of Initial Planetesimal Sizes}

We now examine results for an ensemble of calculations where most of the initial
mass is concentrated in the largest planetesimals, $N_c \propto r_{ik}^{\qprime}$
with \qprime\ = 0.5.  Because protoplanet growth is fairly independent of the
surface density relation, all calculations assume an initial 
$\Sigma = 15 x_m (\mstar / 0.5 \msun) (a /{\rm 1~AU})^{-3/2}$~\gcms. For simplicity, all calculations adopt
the $f_w$ fragmentation parameters. The results in \S3.1 demonstrate that disks 
composed of strong planetesimals yield protoplanets with maximum radii roughly 
10\% larger than disks composed of weak planetesimals 
\citep[see also][]{kb2010,kb2012}. To understand how \rmax\ depends on the initial 
planetesimal size, we consider disks with $r_0$ = 1--1000~km. 

Compared to calculations with 1~m to 1~km planetesimals, we expect several clear 
differences in the timing and maximum radius of protoplanets. With all of the 
mass concentrated in large planetesimals, damping from gas drag is much less 
effective at countering stirring by dynamical friction and viscous stirring. 
Weaker damping yields larger random velocities for small planetesimals. With 
less mass in small planetesimals, dynamical friction cannot brake the 
velocities of the large planetesimals.  Less braking implies larger random 
velocities for the larger planetesimals. Larger velocities lead to longer 
relaxation times and smaller gravitational focusing factors.  Thus, we predict 
large protoplanets form on much longer time scales.

Concentrating the mass in the largest protoplanets also limits fragmentation
and the collisional cascade. Longer periods of orderly growth and less dramatic
runaway growth allow smaller planetesimals to grow larger. Larger planetesimals
are harder to fragment. With less fragmentation, the largest protoplanets accrete 
from a more massive reservoir and can grow larger. For calculations with 
$r_0 \gtrsim$ 3~km, disks have more and more material in very large objects
that are very hard to fragment.  Thus, we predict more massive protoplanets
for calculations with $r_0$ = 1~km and much more massive protoplanets for 
larger $r_0$.

For our typical starting conditions with $e_0 = 10^{-4}$, viscous stirring
often plays a larger role when the calculations begin. For planetesimals with 
$r_0 \gtrsim$ 3~km, planetesimal velocities are in the shear regime, where 
dynamical friction is ineffective \citep{oht2002}. Although growth in the
shear regime is rapid \citep[e.g.,][]{gold2004}, the stirring time scale is
much shorter than the growth time. Thus, viscous stirring rapidly increases
planetesimal velocities and slow, orderly growth begins in the dispersion 
regime \citep[see also][]{kb2012}. 

\begin{figure}
\includegraphics[width=6.5in]{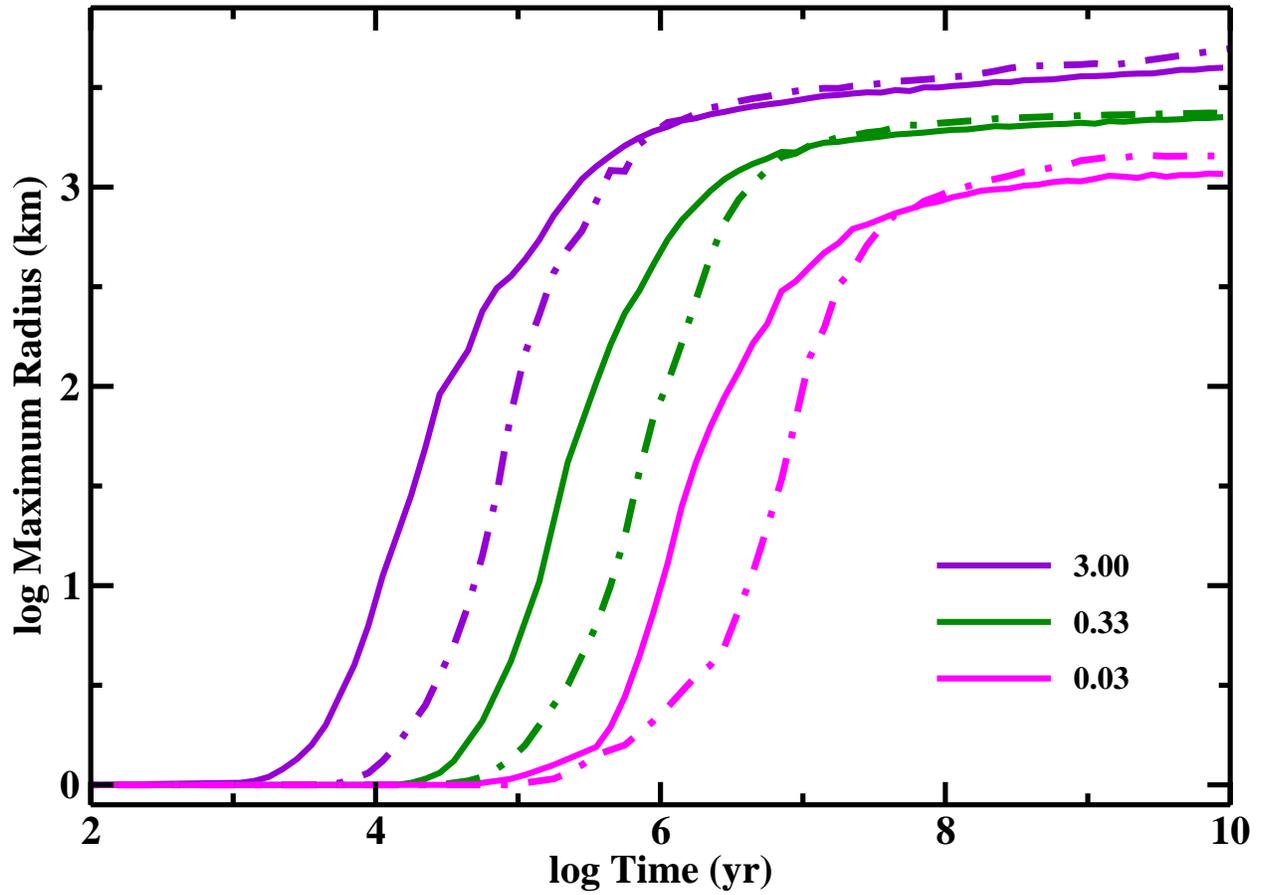}
\vskip 3ex
\caption{%
Evolution of maximum protoplanet radius in disks with $N_c \propto m^{-q^{\prime}}$ 
and $q^\prime$ = 1 (solid lines) or $q^\prime$ = 0.17 (dot-dashed lines) for a 
0.5~\msun\ central star. The legend indicates the initial disk mass \xm\ for each 
curve.  In disks with most of the mass in objects close to the initial maximum size, 
protoplanets grow more slowly but achieve larger masses.
\label{fig: rmax-q}
}
\end{figure}

Fig. \ref{fig: rmax-q} illustrates several of these points for calculations of
disks composed of planetesimals with $r_0$ = 1~km surrounding a 0.5~\msun\ central 
star. When 
\qprime\ = 3 (solid lines in the Fig.), runaway growth begins at 
$\sim 10^4$~yr for \xm\ = 3 (solid violet line), at $\sim 10^5$~yr 
for \xm\ = 0.33 (solid green line), and at $\sim 10^6$~yr for \xm\ = 0.03 
(solid magenta line). When most of the mass is concentrated initially 
in large planetesimals (\qprime\ = 0.5, dot-dashed lines), runaway 
growth begins at $\sim 3 \times 10^4$~yr (\xm\ = 3) to 
$\sim 3 \times 10^6$~yr (\xm\ = 0.03).  Despite the later start, calculations
with \qprime\ = 0.5 result in 5\% larger protoplanets at $t \gtrsim$ 100~Myr. 

\begin{figure}
\includegraphics[width=6.5in]{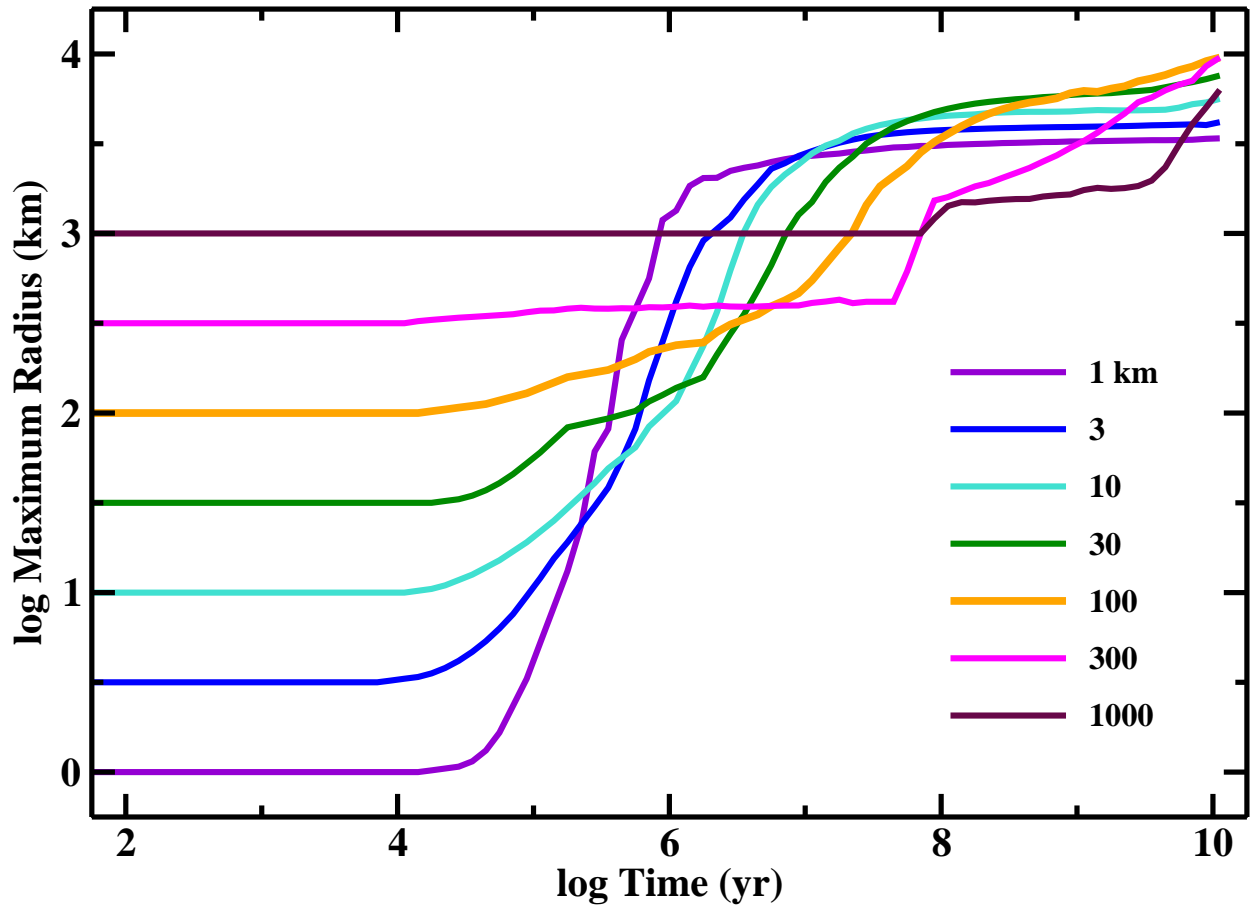}
\vskip 3ex
\caption{%
Evolution of maximum planet radius in disks with \xm\ = 1, $N_c \propto m^{-0.17}$, 
and various initial $r_0$ (as indicated in the legend) around a 0.5 \msun\ star. At late
times (t $\gtrsim$ 1~Gyr), the largest planets form in disks with large planetesimals.
\label{fig: rmax-r0-0p5}
}
\end{figure}

Fig. \ref{fig: rmax-r0-0p5} shows that more massive protoplanets result from 
initially more massive planetesimals. When most of the mass is concentrated
in the largest planetesimals, the growth time initially scales with the size 
of the largest particle size 
\citep[eq. \ref{eq: t-coll}; see also][]{gold2004,ormel2010c}. Thus, disks 
composed of small planetesimals achieve runaway growth faster than disks
composed of large planetesimals.  However, runaway growth with smaller
planetesimals leaves more material in objects that are easier to fragment.
Thus, growth with 30--300~km planetesimals produces much larger protoplanets
than growth with 1--10~km planetesimals \citep[see also][]{kb2010}. 

Calculations with most of the initial mass in 1000~km planetesimals
are an exception to this trend (Fig. \ref{fig: rmax-r0-0p5}). When
\xm\ = 1, it takes more than 100~Myr for collisions to produce objects
somewhat larger than 1000~km.  This slow growth phase lasts nearly
5~Gyr, when `runaway growth' yields objects larger than
3000~km. However, this runaway is still slow: it takes another 5~Gyr
for objects to grow to 5000~km and another few Gyr to reach
$10^4$~km. Thus, massive planets reach their maximum mass at times,
$\gtrsim$ 15~Gyr, larger than the age of the universe.

\begin{figure}
\includegraphics[width=6.5in]{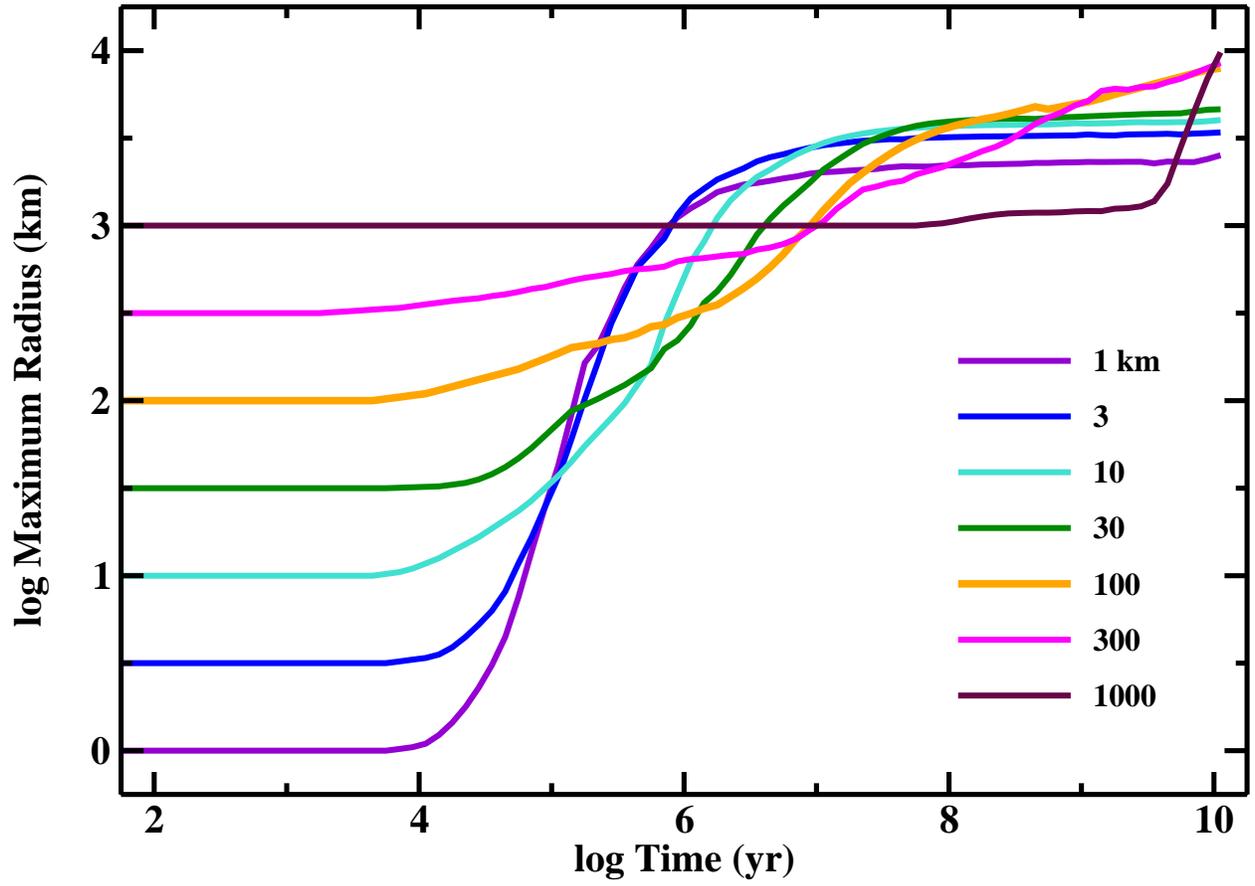}
\vskip 3ex
\caption{%
As in Fig. \ref{fig: rmax-r0-0p5} for a 0.3 \msun\ star.
\label{fig: rmax-r0-0p3}
}
\end{figure}

\begin{figure}
\includegraphics[width=6.5in]{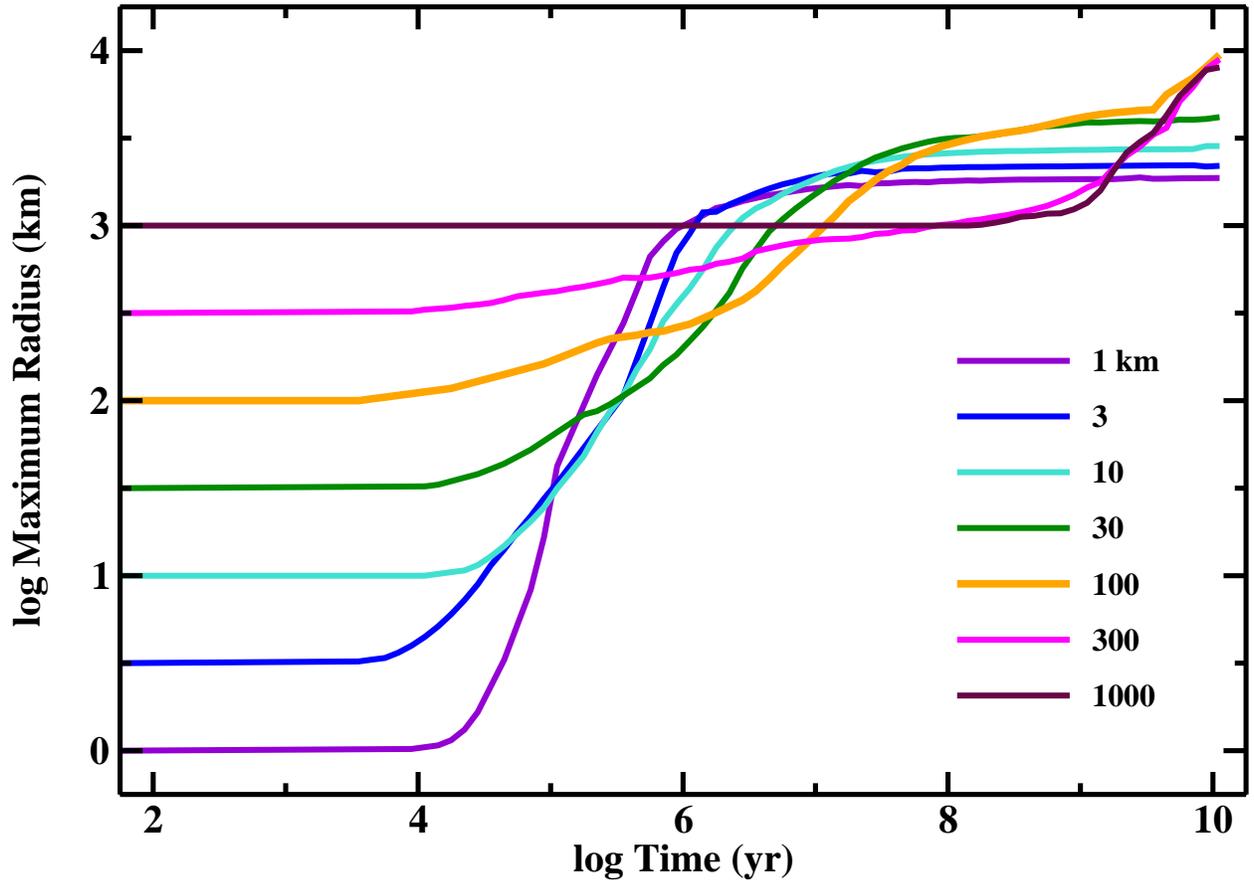}
\vskip 3ex
\caption{%
As in Fig. \ref{fig: rmax-r0-0p5} for a 0.1 \msun\ star.
\label{fig: rmax-r0-0p1}
}
\end{figure}

Figs. \ref{fig: rmax-r0-0p3}--\ref{fig: rmax-r0-0p1} repeat plots of the 
evolution of
\rmax\ with time for calculations with 0.3~\msun\ and 0.1~\msun\ central stars.
These plots show the same trends. Calculations with smaller planetesimals 
reach runaway growth sooner, but collisional cascades remove more material
from the disk before large protoplanets can reach their maximum sizes. Thus,
disks with 30--300~km planetesimals produce the largest protoplanets.

\begin{figure}
\includegraphics[width=6.5in]{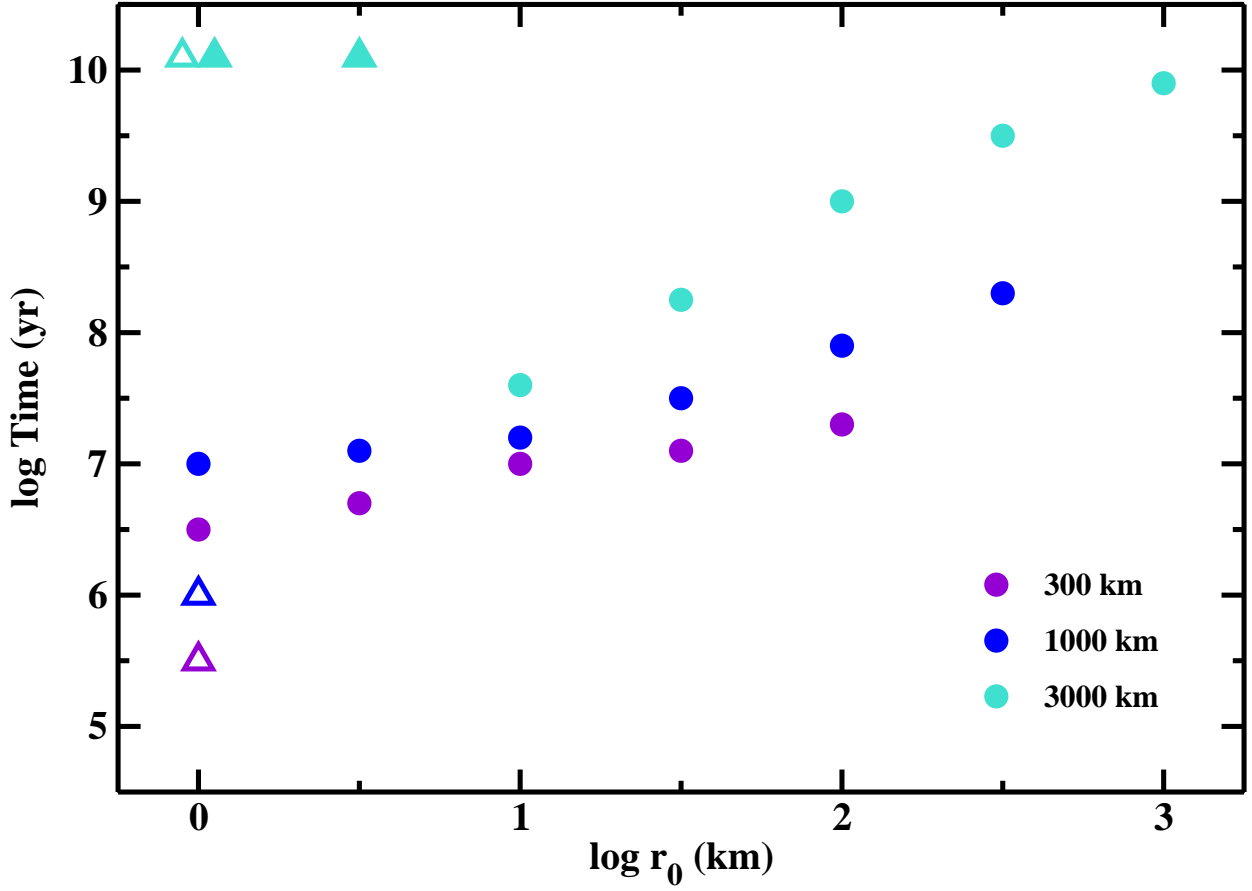}
\vskip 3ex
\caption{%
Evolution time required to achieve 
\rmax\ = 300~km (violet symbols),
\rmax\ = 1000~km (blue symbols), and
\rmax\ = 3000~km (turquoise symbols)
as a function of initial planetesimal size $r_0$ for
calculations with a 0.5 \msun\ central star.
Filled symbols illustrate results for calculations
with weak planetesimals and most of the initial 
mass in planetesimals with $r \approx$ \r0 
($N_c \propto m^{-0.17}$).  
Open symbols show results for calculations with
weak planetesimals and $N_c \propto m^{-1}$.  
The triangles in the upper left of the plot indicate 
lower limits.
\label{fig: time-r0}
}
\end{figure}

To illustrate these conclusions in more detail, Fig.~\ref{fig: time-r0} shows 
the time required to produce 300~km ($t_{300}$), 1000~km ($t_{1000}$), and
3000~km ($t_{3000}$) objects as a function of the initial planetesimal size 
$r_0$ (see Table 5).  When the initial mass in planetesimals is spread over 
objects with radii of 1~m to 1~km,
growth rapidly produces 300~km and 1000~km objects (open triangles). Growth with
most of the mass in 1~km planetesimals takes roughly three times longer. In both
cases, the collisional cascade prevents protoplanets from reaching 3000~km 
(symbols at log time = 10.1).  As $r_0$ increases, it takes longer and longer 
to produce 300--1000~km protoplanets. Aside from the $r_0$ = 1-3~km models, all
calculations with most of the mass in large planetesimals produce 3000~km 
objects.  

\begin{figure}
\includegraphics[width=6.5in]{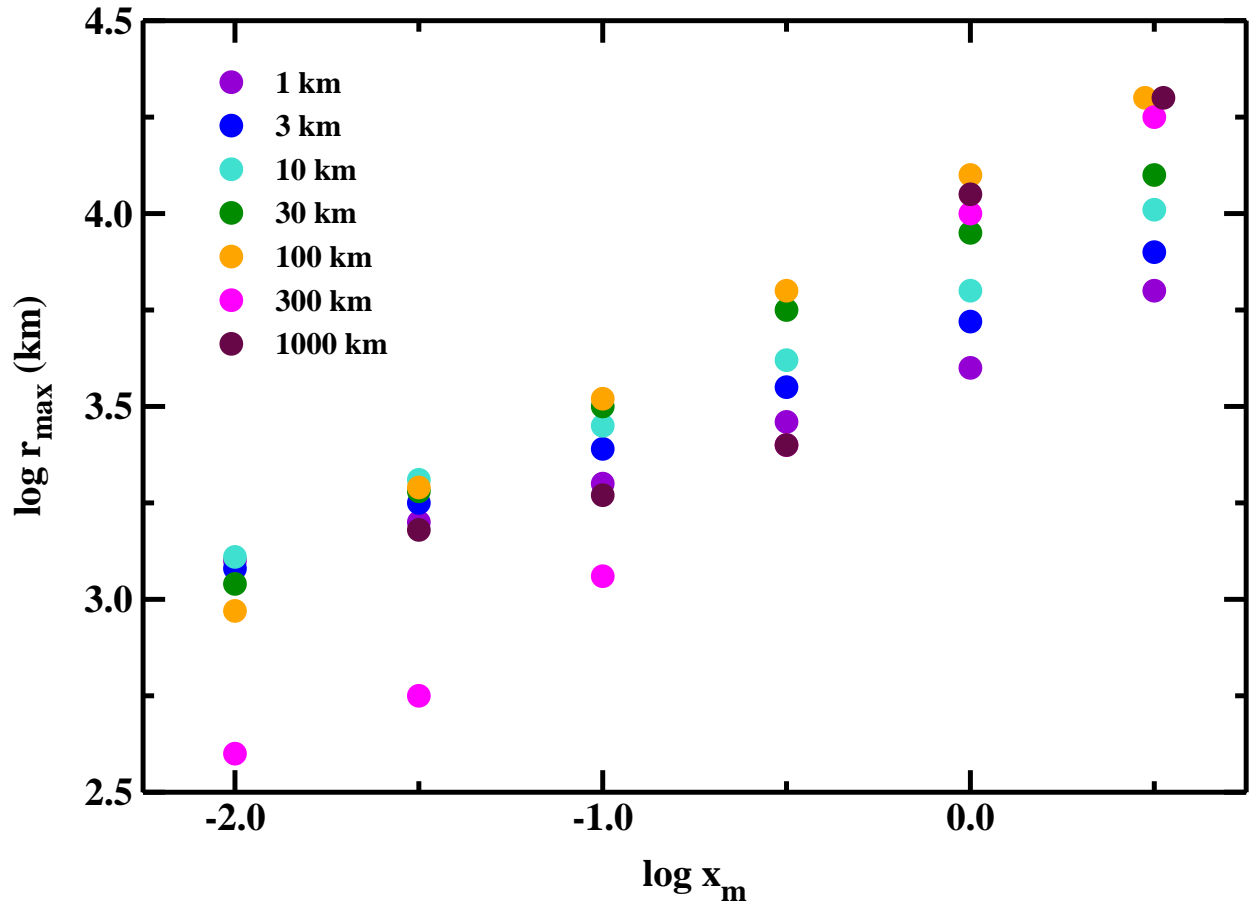}
\vskip 3ex
\caption{%
The size of the largest object as a function of initial
disk mass for calculations with a 0.5 \msun\ central star,
strong planetesimals, and most of the mass initially in 
the largest planetesimals. The legend indicates $r_0$, the
initial size of the largest planetesimal.
\label{fig: rmax-xm}
}
\end{figure}

Although larger planetesimals grow slowly, they often produce much larger 
protoplanets (Fig.~\ref{fig: rmax-xm}). At small disk masses (\xm\ $\lesssim$ 0.1), 
the long time scales leading to runaway growth prevent the formation of massive 
protoplanets. Smaller planetesimals then produce larger protoplanets. For larger 
disk masses (\xm\ $\gtrsim$ 0.1), however, larger planetesimals yield much more 
massive protoplanets. The time scales for large planetesimals to reach runaway growth 
are then long but still shorter than a Hubble time. With reduced losses from the 
collisional cascade, these disks have more than enough time to produce 
super-Earth-type planets in 5--10~Gyr.

\begin{figure}
\includegraphics[width=6.5in]{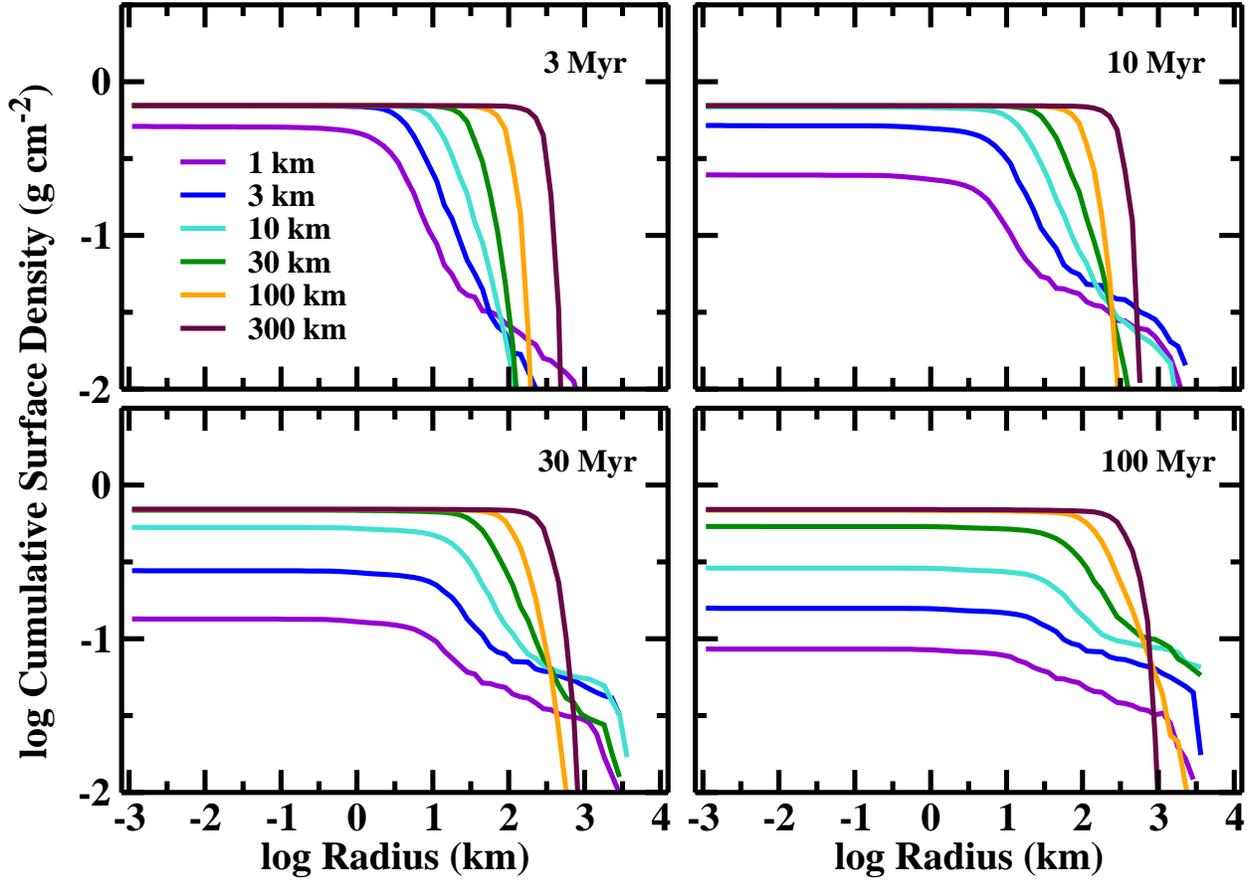}
\vskip 3ex
\caption{%
Evolution of the size distribution at 7--8 AU for models with \xm\ = 1, 
$N_c \propto m^{-0.17}$, and various $r_0$ (as indicated in the legend in the 
upper left panel).  At 3 Myr (upper left panel), disks with $r_0$ = 1~km
produce 1000~km objects and lose roughly 10\% of the initial mass to debris. 
By 10 Myr (upper right panel), disks with $r_0$ = 3~km produce larger
objects and also begin to lose mass in the collisional cascade. At later times,
collisional cascades start to remove mass from disks composed of 10~km objects
(30 Myr, lower left panel) and 30~km objects (100 Myr, lower right panel). 
At 100 Myr, calculations with $r_0$ = 100~km and 300~km commence producing 
larger objects.
\label{fig: sd1}
}
\end{figure}

\begin{figure}
\includegraphics[width=6.5in]{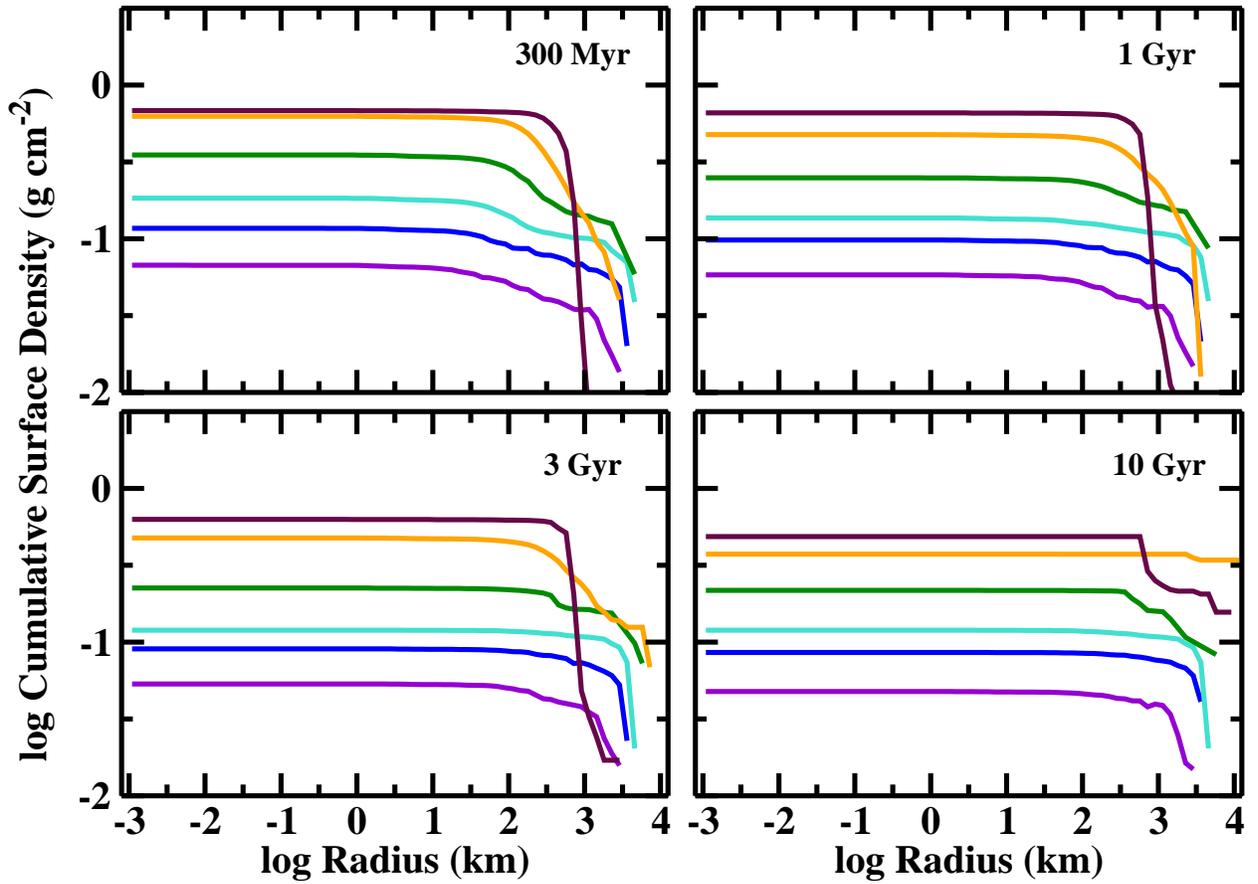}
\vskip 3ex
\caption{%
As in Fig. \ref{fig: sd1} for later evolution times. From 300 Myr to 10 Gyr,
disks with $r_0$ = 1--30~km have little growth in the largest objects and
lose more and more mass to debris. Disks with $r_0$ = 100--300~km lose little
mass to debris. These disks produce super-Earths with $r \gtrsim 10^4$~km
at 5--10 Gyr ($r_0$ = 100~km) and at 9--12 Gyr ($r_0$ = 300~km).
\label{fig: sd2}
}
\end{figure}

To conclude this section, Figs.~\ref{fig: sd1}--\ref{fig: sd2} plot the evolution
of the median size distribution for calculations with $r_0$ = 1--1000~km. The first 
figure shows the evolution for the first 100~Myr; the second figure shows the 
evolution for 300~Myr to 10~Gyr. 

In Fig.~\ref{fig: sd1}, the upper left panel shows the rapid growth of an ensemble
of 1~km planetesimals relative to other starting sizes. At 3~Myr, some objects 
have already reached radii of 1000~km. However, stirring by these large objects
has also started the collisional cascade: the cumulative surface density is roughly
10\% smaller than the starting $\Sigma$. As time proceeds, the largest objects in
this calculation continue to grow; the surface density continues to fall. After
300 Myr (lower right panel), the collisional cascade has removed roughly 90\% of 
the initial mass in solid material.

For ensembles of larger planetesimals, the rate of growth and the loss of material 
from the collisional cascade correlate with initial planetesimal size. At 10~Myr,
calculations with $r_0$ = 3~km contain the largest objects. By 100~Myr, calculations
with $r_0$ = 10--30~km contain the largest objects. Within this sequence of \r0, 
the fraction of material lost to fragmentation declines from 90\% ($r_0$ = 1~km)
to 80\% (3~km) to 35\% (10~km) to 10\% (30~km). Lower fragmentation rates allow
calculations with larger $r_0$ to produce larger protoplanets.

In Fig.~\ref{fig: sd2}, the evolution from 300~Myr to 10~Gyr shows similar trends. 
After 10~Gyr, calculations with $r_0$ = 1~km have lost almost 95\% of their initial mass;
simulations with 300--1000~km planetesimals have lost less than 5\%. As a result, disks 
initially composed of large planetesimals produce much larger protoplanets.

\subsection{Summary of Coagulation Calculations}

The coagulation calculations in \S3.1 and \S3.2 demonstrate that collisional
evolution of a disk of planetesimals leads to the production of planets ranging
in mass from Pluto to roughly ten Earth masses. From the full suite of calculations,
we infer five specific conclusions.

\begin{itemize}

\item Results from coagulation calculations match analytic theory.  For objects 
with radii 100--1000~km, the growth time is $t \propto x_m^{-\gamma_2} a^{k + 3/2}$ 
where $\gamma_2 \approx$ 1.1 and $k$ is the exponent in the power-law relation 
between surface density and semimajor axis.

\item Outcomes of planet formation are relatively insensitive to the slope of the power 
law relation between the disk surface density and the semimajor axis. For similar total
masses, disks with shallower surface density gradients have less (more) mass in 
the inner (outer) disk than disks with steeper surface density gradients.  
Analytic theory and coagulation calculations demonstrate that the time scale to 
produce protoplanets with radius \rmax\ depends on the local surface density.
Thus, massive protoplanets form more rapidly in the inner regions of disks with
steep surface density gradients and in the outer regions of disks with shallow
surface density gradients. However, the maximum protoplanet radius \rmax\ does
not depend on the surface density gradient.

\item Fragmentation is important for setting the growth time and the maximum 
radius of growing protoplanets with $r \gtrsim$ 300--500~km.  When planetesimals are 
small and relatively weak, protoplanets with $r \gtrsim$ 1000~km take longer to form
and grow to smaller masses. For calculations starting with 1--3~km planetesimals
and `standard' fragmentation parameters, \rmax\ $\approx$ 3000~km. These systems
probably produce copious amounts of dust \citep[see also][]{kb2008, kb2010}.

\item The initial sizes of planetesimals have a major impact on the growth of 
protoplanets. When most of the solid mass is initially in large planetesimals with 
radius \r0, protoplanets grow more slowly with increasing \r0. As $r_0$ increases,
however, fragmentation has a smaller and smaller impact on protoplanet growth. Thus,
calculations with larger $r_0$ yield larger protoplanets. 

\item The initial size distribution of planetesimals also plays a critical role in 
outcomes of planet formation. When $r_0$ $\gtrsim$ 0.1~km and the initial size
distribution is broad, a balance between viscous stirring and gas drag enables
large gravitational focusing factors and promotes rapid growth of protoplanets.
When the size distribution is narrow, gas drag is less effective. Growth is slower.

\end{itemize}

There are two main sources of uncertainty in these conclusions.

\begin{itemize}

\item In disks composed of large planetesimals, Earth-mass protoplanets may eject 
leftover planetesimals before reaching super-Earth masses. As they grow, protoplanets 
stir up much smaller objects within several Hill radii of their orbits to velocities 
comparable to their escape velocity, $\sim$ 10~\kms\ for an icy Earth-mass planet. At 
the inner edges of the disks considered here, the escape velocity from the star is also 
$\sim$ 10~\kms.  Thus, Earth-mass planets clear their orbits of smaller objects and
eject these objects from the planetary system.  However, as the mass of the protoplanet 
approaches an Earth-mass, small objects with radii less than 1000~km contain less than 
$\sim$ 5\% of the mass of the protoplanet. Thus, scattering probably has little impact 
on the final mass of the protoplanet.

\item Merger rates of sub-Earth mass protoplanets into Earth mass or larger protoplanets
rely on the particle-in-a-box cross-section instead of direct $N$-body calculations.
As more and more mass is concentrated into a few large objects, the particle-in-a-box
cross-section provides a poorer and poorer representation of the likely collision rate.
Several test simulations using the $N$-body component of {\it Orchestra} suggest that
mergers among sub-Earth mass protoplanets occur over a somewhat broader range of time 
scales than suggested by pure coagulation calculations. However, mergers still occur on
roughly 1~Gyr time scales. Thus, more sophisticated calculations are unlikely to change 
our general conclusions.

\end{itemize}

Despite some uncertainties, our results lead to several broad conclusions for the outcomes 
of protoplanet growth around low mass stars. 

\begin{enumerate}

\item In systems with short gas depletion times, disks composed primarily of large 
planetesimals with $r_0$ $\gtrsim$ 10~km are unlikely to form gas giant planets.  
In the core accretion theory, formation of gas giant planets requires the growth 
of a multi-Earth mass icy core before the dissipation of the gaseous disk 
\citep[e.g.,][]{pollack1996,liss2009}.  With a gas dissipation time scale of 
$\sim$ 1~Myr, icy cores must form in $\lesssim$ 1~Myr.  In disks of large 
planetesimals, it takes much longer than 1~Myr to form Earth-mass icy cores.

\item Disks composed primarily of large planetesimals can produce icy super-Earths.
Although super-Earth formation occurs at 1--10~Gyr, the long main sequence lifetimes 
of 0.1--0.5 \msun\ stars guarantee that massive disks will form several icy 
super-Earths at distances of several AU from their host stars. 

\item In disks composed of small planetesimals with $r_0$ $\lesssim$ 1--3~km,
1000--2000~km icy cores grow in less than 1~Myr. With the gas surface density 
still large, the gas can damp the velocities of small particles, preventing 
the collisional cascade \citep{raf2004,koba2010b}.  If the cascade produces 
large numbers of small particles with radii of $\sim$ 1 cm, low mass icy 
cores rapidly accrete them and evolve into Earth mass or larger objects 
\citep{raf2004,kb2009,ida2010,ormel2010a,ormel2010b,bk2011a,lamb2012}. 
However, gas drag may halt the cascade at larger sizes, 1--10~m, where 
accretion by large objects is less effective \citep[e.g.,][]{koba2010b,koba2011}. 
Thus, these systems seem unlikely sources of super-Earths or gas giant planets.

\end{enumerate}

Constructing observational tests from these predictions requires one additional
ingredient, $a(t)$ for newly-formed protoplanets. With current technology, only
direct imaging and microlensing observations can detect super-Earths over the 
range of semimajor axes considered in our calculations. If planets drift inward
significantly, however, then radial velocity and transit observations can detect 
them.  For massive planets, strong gravitational interactions among closely-packed 
planets at large $a$ often yield one or more planets orbiting much closer to the 
central star \citep[e.g.,][]{rasio1996,juric2008,mann2010,naga2011}.  Because large ensembles 
of scattering calculations are cpu intensive, we plan a separate study of this issue.
Less dramatically, radial migration through the circumstellar disk can deposit planets 
of all masses close to the central star \citep[e.g.,][]{ida2008a,mord2009}.  Thus,
we explore whether migration can transport super-Earths formed at late times from 
$a \approx$ 2--10~AU to $a \ll$ 1 AU.

\section{MIGRATION}
\label{sec: mig}

To examine whether planets formed at 2--10~AU migrate inwards, we consider super-Earths
formed on time scales of $\sim$ 100~Myr to 1~Gyr in a protostellar disk.
Although migration of super-Earths through a young gaseous disk is often 
efficient \citep{linpap1986,ward1997,paarde2010}, there is no disk gas 
remaining after $\sim$ 100~Myr. Thus, we consider the radial motion of 
super-Earths through a remnant disk of solid debris 
\citep{ward1997,kirsh2009,bk2011b,ormel2012}.  

\subsection{Migration Through a Planetesimal Disk}

Migration through a disk of solid material depends on the mass of the planet and 
the local properties of the planet.  Close to the planet, the Hill sphere defines 
a region where the gravity of the planet roughly balances the gravity of the 
central star.  If $m_p$ is the mass of the planet and $a_p$ is its orbital 
semimajor axis, the Hill radius is
\begin{equation}
\rhill = \left ( { m_p \over 3 \mstar } \right )^{1/3} a_p ~ .
\label{eq: rhill}
\end{equation}
For low mass stars,
$\rhill \approx 0.015 ~ (m_p / \mearth)^{1/3} ~ (\mstar / 0.3~\msun)^{-1/3} a_p$.

Interactions between a planet and nearby planetesimals set the direction and pace
of migration.  A large planet tries to clear its orbit by scattering smaller planetesimals.  
Scattering among planetesimals tries to fill the orbit of the planet. When the planet 
dominates, it reduces the surface density of planetesimals along its orbit and increases 
the surface density of planetesimals on either side of its orbit. Because the surface 
density enhancements are not axisymmetric, the planet feels a torque from the planetesimals.  
If the sum of all of the torques does not vanish, the large objects migrate radially 
inward or outward \citep[e.g.,][]{gold1982,ward1997,ida2000b,kirsh2009,ormel2012}.  

When scattering overcomes viscous spreading, planets open gaps in the radial distribution 
of planetesimals. For convenience, we define the Hill radius necessary for a planet to 
open up a gap in a cold disk of solid planetesimals where the vertical scale height $H_z$
is smaller than the Hill radius.  Adopting typical conditions in disks around low mass 
stars, this radius is \citep[e.g.,][]{raf2001,bk2013}
\begin{equation}
\rgap \approx 0.03 ~ x_m^{1/3} 
\left ( { R \over {\rm 1~km} } \right )^{-1/3}
\left ( { a \over {\rm 10~AU} } \right ) 
\left ( { 0.3 \msun \over \mstar\ } \right )^{1/3} ~ {\rm AU} ~ .
\label{eq: rgap}
\end{equation}
Planets with \rhill\ $\approx$ 0.03~AU at 10~AU have 
$m_p \approx$ 0.008~\mearth\ (eq. [\ref{eq: rhill}]). Thus, low mass planets open up gaps
in disks of planetesimals.

For planets with $\rhill \gtrsim R_{gap}$, there are two possible modes of migration
\citep[e.g.,][]{ida2000b,kirsh2009,bk2011b,bk2013}. If the planet can clear a gap in its corotation zone and 
migrate across this gap in one synodic period, the planet undergoes fast migration. This
mode is similar to type III migration through a gaseous disk \citep{masset2003}.  
Planets with Hill radii in the range 
\rgap\ $\lesssim \rhill \lesssim$ \rfast\ ($m_p \approx$ 0.01--0.3 \mearth) satisfy this 
condition, where 
\begin{equation}
\rfast \approx 0.1 ~ x_m^{1/2} ~ 
\left ( { a \over 10~{\rm AU} } \right )^{3/2}
\left ( { 0.3 \msun \over \mstar} \right )^{1/3} ~ {\rm AU} ~ .
\label{eq: rfast}
\end{equation}
As in type II gaseous migration \citep[e.g.,][]{syer1995}{}, planets with 
$\rhill \gtrsim \rfast$ migrate through the disk more slowly.

For all three modes of migration through a disk of planetesimals, the pace of inward or
outward migration is more leisurely than migration through gas \citep{bk2011b,bk2013}.  
For planets with $\rhill < \rgap$ in a cold disk where $\Sigma \propto a^{-3/2}$ (see
eq. [\ref{eq:sigma-dust}]) and where $e_0$ for the background planetesimals is small, 
analytic results yield
\begin{equation}
\dot{a}_{slow} 
%\approx 
%-1.3 \times 10^{-9} x_m \left ( { m_p \over \mearth\ } \right ) \left ( { 0.3 \msun\ \over \mstar } \right )^{1/2} \left ( { a \over {\rm 10~AU} } \right)^{1/2} ~ {\rm AU~yr^{-1}} ~ .
\lesssim
-10^{-7} x_m \left ( { m_p \over \mearth\ } \right ) \left ( { 0.3 \msun\ \over \mstar } \right )^{1/2} \left ( { a \over {\rm 10~AU} } \right)^{1/2} ~ {\rm AU~yr^{-1}} ~ .
\label{eq: dadt-type1}
\end{equation}
Planets with $\rgap \simless \rhill \simless \rfast$ migrate rapidly in the fast mode. This
rate is independent of the mass of the planet:
\begin{equation}
\dot{a}_{fast} \approx 
%\pm 10^{-4} x_m \left ( { 0.3 \msun\ \over \mstar } \right )^{1/2} \left ( { a \over {\rm 10~AU} } \right )^{1/2} ~ {\rm AU~yr^{-1}} ~ .
\pm 10^{-3} x_m \left ( { 0.3 \msun\ \over \mstar } \right )^{1/2} \left ( { a \over {\rm 10~AU} } \right )^{1/2} ~ {\rm AU~yr^{-1}} ~ .
\label{eq: dadt-type3}
\end{equation}
For planets capable of opening up a gap in the disk ($\rhill \gtrsim \rgap$), the expected migration rate is:
\begin{equation}
%\dot{a}_{gap} \approx -4.4 \times 10^{-6} x_m \left ( { m_p \over \mearth\ } \right )^{1/3} \left ( { 0.3 \msun\ \over \mstar } \right )^{1/2} \left ( { a \over {\rm 10~AU} } \right)^{1/2} ~ {\rm AU~yr^{-1}} ~ .
\dot{a}_{gap} \approx -10^{-6} x_m \left ( { m_p \over \mearth\ } \right )^{1/3} \left ( { 0.3 \msun\ \over \mstar } \right )^{1/6} \left ( { a \over {\rm 10~AU} } \right)^{1/2} ~ {\rm AU~yr^{-1}} ~ .
\label{eq: dadt-type2}
\end{equation}
Planet migrate rapidly (i) when the disk is more massive, (ii) when the planet is more massive, and 
(iii) when the planet is farther away from the central star. 

In a warm disk, migration rates depend on the eccentricity of planetesimals 
\citep[e.g.,][]{ida2000b,kirsh2009,bk2011b}. Using an elegant analytic approach, \citet{ormel2012} 
derive rates from encounters with distant and nearby planetesimals. Their results indicate 
$\dot{a} \propto e^{-2}$. For an Earth-mass planet with $e \approx$ 0.02 at 10~AU around a 
0.3~\msun\ star, their migration rates are comparable to our analytic rates. 

The analytic results for planets embedded in cold planetesimal disks have several consequences for 
growing icy planets around low mass stars.

\begin{itemize}

\item Low mass planets with $m_p \lesssim 0.01 \mearth$ cannot migrate very far through the disk. 
For disks with $x_m \approx$ 1 in leftover planetesimals, the nominal rates are roughly 1~AU per Gyr. 
Once large planets capable of migrating form, likely surface densities of leftover planetesimals 
in the disk are much smaller.

\item Although intermediate mass planets with $m_p \approx$ 0.01-0.3 \mearth\ can migrate rapidly,
fast mode migration is probably rare. Oligarchic growth usually produces many planets in this
mass range. These oligarchs rapidly stir up leftover planetesimals along their orbits. As they 
migrate, they leave behind wakes of highly stirred planetesimals. Oligarchs cannot migrate 
through these wakes \citep{bk2011b}. Thus, fast migration generally stalls until many oligarchs 
merge into a few massive oligarchs. After these mergers, the radii of massive oligarchs exceed 
\rfast. These oligarchs then open a gap in the disk and migrate more slowly.

\item Massive oligarchs with $m_p \gtrsim 0.3 \mearth$ migrate at typical rates of 1 AU Myr$^{-1}$. 
Massive oligarchs can migrate through the wakes of other oligarchs and can drift close to the 
central star. As the planet migrates inward, rates decline. Continued stirring and depletion of 
the disk by fragmentation also slows migration \citep{bk2011b}.  Even if rates slow by 2--3 
orders of magnitude, the long lifetimes of 0.1--0.5 \msun\ stars suggest that some icy planets 
might migrate from $\sim$ 10~AU to $\sim$ 0.1~AU. 

\end{itemize}

These considerations indicate that massive planets probably migrate through the leftover
planetesimal disk on time scales of 1~Myr to 1~Gyr. Because the analytic rates rely on 
the properties of cold disks instead of self-consistent disk models, testing this conclusion 
requires numerical calculations of planet migration through plantesimal disks. To explore
likely outcomes, we now discuss representative calculations of planets migrating through 
disks of leftover planetesimals.

\subsection{Numerical Calculation of Migration Through a Planetesimal Disk}

To examine migration through a protostellar disk in more detail, we consider a suite of
numerical simulations of a single planet embedded in a planetesimal disk. The simulations
explore the behavior of icy planets ($R \approx$ 1~\rearth, $m \approx$ 0.25~\mearth\ and 
$R \approx$ 2~\rearth, $m \approx$ 2.5 \mearth) at semimajor axes $a$ = 1, 3, and 10 AU
in orbit around 0.1, 0.3, and 0.5 \msun\ stars. To derive the range of plausible migration 
rates, we consider disks where the surface density scale factor is \xm\ = 0.01, 0.1, and 1
(see eq. [\ref{eq:sigma-dust}]). 

To perform the calculations, we use the $n$-body component of
\orch. We model the planetesimal disk as a swarm of $10^5 - 10^6$
massless tracer particles with a radial extent of 0.75$a_0$ to
2.0$a_0$, where $a_0$ is the initial semimajor axis of the embedded 
planet.  To enable extensive simulations with large number of tracers, 
we calculate the migration of the planet from the total change in angular
 momentum of the tracers \citep{bk2013}. Without an embedded planet, 
long integrations of the tracers conserve angular momentum to machine 
accuracy.  Migration rates derived from the angular momentum of massless 
tracers agree very well with rates derived from more cpu intensive 
calculations with massive swarm particles \citep{bk2013}.

\begin{figure}
\includegraphics[width=6.5in]{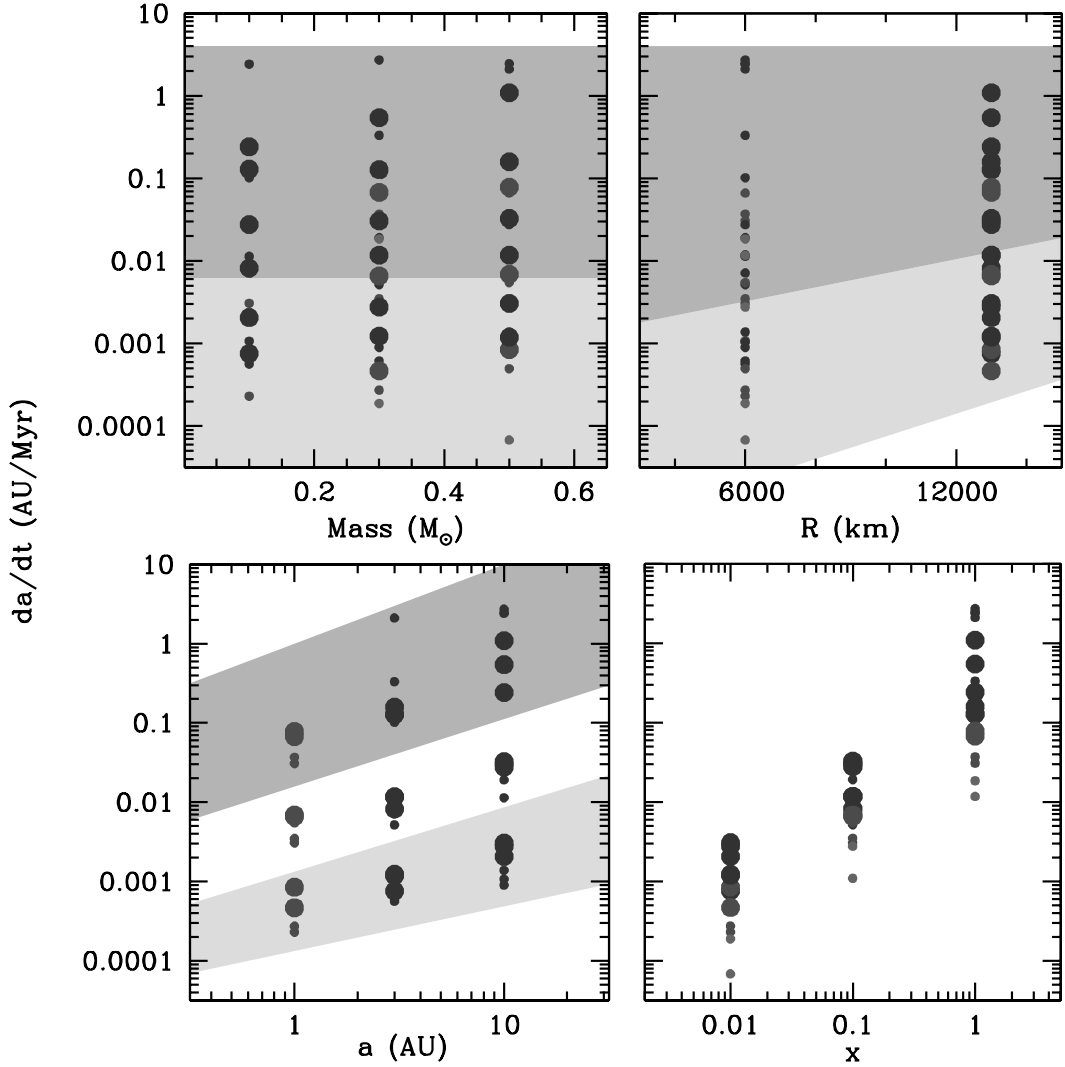}
\vskip 3ex
\caption{%
Migration rates for Earth-mass planets at semimajor axes $a$ = 1, 3, and 10 AU 
around 0.1--0.5 \msun\ stars. Small (large) filled circles indicate results for 
planets with $m$ = 0.25 \mearth\ (2.5 \mearth).  In the lower left panel, shaded 
regions show results as a function of scaled surface density: \xm\ = 0.01 (light 
shading), \xm\ = 0.1 (no shading), and \xm\ = 1.0 (heavy shading).  In the upper 
two panels, shading indicates results for \xm\ $\lesssim$ 0.1 (light shading) 
and \xm\ $\gtrsim$ 0.1 (heavy shading).
\label{fig: migrate}
}
\end{figure}

Figure \ref{fig: migrate} summarizes our results. In the lower right
panel, the migration rate depends on the local disk surface density,
with $\dot{a} \approx -1.5 \xm$~AU~Myr$^{-1}$ for sub-Earth mass and
super-Earth mass planets orbiting 0.1--0.5 \msun\ stars. In all disks,
a migrating planet scatters lower mass planetesimals along its
orbit. Angular momentum and energy exchange during a scattering event
produces migration. More massive disks supply more planetesimals to
the planet's Hill sphere. Thus, planets migrate more rapidly in more
massive disks.

Migration is also sensitive to the planet's semimajor axis
(Fig. \ref{fig: migrate}, lower left panel). The rates scale as
$\dot{a} \approx -1.5 \xm\ (a / {\rm 10~AU})$ AU~Myr$^{-1}$.  The
variation of disk mass with $a$ produces most of the change of
$\dot{a}$ with $a$.  Disks with $\Sigma \propto a^{-1}$ have more mass
at large $a$ than at small $a$. Thus, planets at larger $a$ encounter
and scatter more planetesimals along their orbits than planets at
smaller $a$.

Trends in the migration rate with planet mass (Fig. \ref{fig:
  migrate}, upper right panel) demonstrate that more massive planets
migrate more rapidly. For planets in low mass disks, the rate scales
with $m^\beta$, with $\beta \sim 1/4$--1/2. In more massive disks, the
migration rates scale more weakly with the mass of the planet.

Finally, there is no strong trend in the migration rate with stellar
mass (Fig. \ref{fig: migrate}, upper left panel).  Much of this formal 
insensitivity to $M_\star$ comes from the way in which we define $x_m$; 
a constant value of $x_m$ corresponds to a constant ratio of $\Sigma/M_\star$. 
Still, over a broad range of disk masses and semimajor axes, the drift 
rate for a planet of mass $m$ varies little with stellar mass for 
0.1--0.5 \msun\ stars

\begin{figure}
\includegraphics[width=6.5in]{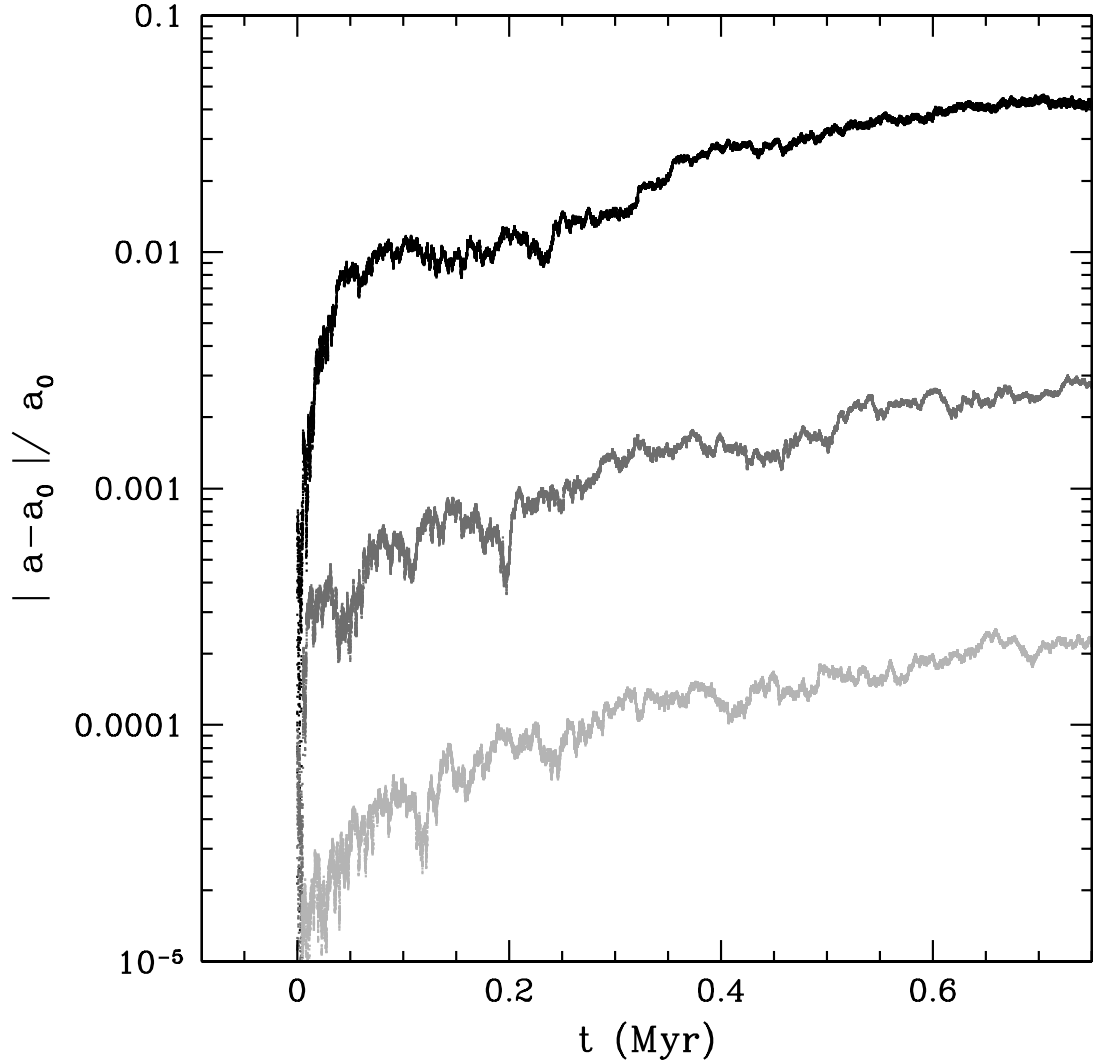}
\vskip 3ex
\caption{%
The radial drift of a planet in 
a planetesimal disk around a 0.3~$\msun$ star.
The curves show the fractional change in semimajor axis $a$,
relative to the starting value $a_0 = 10$~AU, for a planet of
radius $r = 13,000$~km. The curves each correspond to a different
disk mass; $x_m = $0.01, 0.1 and 1 for the light-, medium- and dark-shaded
curves, respectively. In all cases, the drift is inward.
\label{fig: drift}
}
\end{figure}

Figure \ref{fig: drift} emphasizes the importance of disk mass to
radial migration. There is a clear linear dependence of the drift rate 
on $x_m$.  The Figure also illustrates the `choppiness' of migration 
in planetesimal disks.  Even though the mass of the migrating planet 
is between $10^3$ and $10^6$ times larger than the mass of individual 
``planetesimals'' in the disk, the drift rate changes due to 
discreteness noise. The migrating planet simply does not encounter
planetesimals at a constant rate.  We expect this phenomenon to 
persist in more realistic simulations of growing planets, where the 
distribution of the smaller objects is broad in mass and patchy in
space.

In general, the numerical results agree with analytic expectation. The
basic migration rate of 0.1--1~AU~Myr$^{-1}$ derived from the
simulations agrees well with the analytic rate of
0.4--0.5~AU~Myr$^{-1}$ (see eq. [\ref{eq: dadt-type2}]). As expected,
the migration rate scales linearly with disk mass. Although the
numerical migration rates scale with $a$, $m$, and \mstar, the
power-law relations implied by the numerical simulations differ from
the analytic results. In the simulations, migration rates are more
sensitive to $a$ and less sensitive to $m$ or \mstar. We speculate
that these differences result from the way migration proceeds when the
small planetesimals have a vertical scale height larger than the Hill
radius of the planet \citep[see also][]{bk2011b}. Verifying this
speculation requires many more simulations which are beyond the scope
of this paper.

\subsection{Semi-analytic migration model}

Constructing predictive models for the migration of super-Earths through 
a disk of planetesimals around a low mass star requires a hybrid calculation 
covering a broad range of semimajor axes, 0.1--10~AU, which follows 
(i) the growth of planets as a function of semimajor axis with the coagulation 
code and (ii) the migration of planets through the sea of left over planetesimals 
with the $n$-body code. This calculation is computationally expensive.  
To develop an initial set useful predictions, we first explore a simple algorithm.  

Our model assumes a massive planet grown within a sea of small particles with an 
initial surface density scale factor \xm\ = 1. Mergers of planetesimals into a 
planet depletes the surface density. Thus, the scale factor declines with time. 
Planets form more rapidly at smaller $a$ (eq. [\ref{eq: t-coll-fin}]). At a 
given time, the scale factor increases with $a$: the inner disk is more depleted 
than the outer disk. 

To establish a reference for migration in a depleted disk, we begin with migration
in disks with constant \xm.  From Fig. \ref{fig: migrate}, we adopt a simple 
expression for the migration rate:
\begin{equation}
\dot{a} = -1.5 \times 10^{-6} \xm\ \left ( { a \over {\rm 10~AU} } \right ) \left ( {m \over {\rm 2.5~\mearth} } \right )^{1/2} ~ {\rm AU~yr^{-1}} ~ .
\label{eq: adot-sim}
\end{equation}
For the initial conditions, we place a single planet at the inner edge of the disks 
considered in \S3.  Thus $a_0$ = 7 AU for a 0.5~\msun\ star, $a_0$ = 4 AU for a 
0.3~\msun\ star, and $a_0$ = 2.5 AU for a 0.1~\msun\ star.  With \xm\ fixed, it is 
straightforward to integrate the time evolution of $a$ for a planet with mass $m$.

This approach ignores the variation of $\dot{a}$ with the eccentricity of leftover planetesimals.
In \citet{ormel2012}, migration rates for $e \lesssim$ 0.02--0.05 are a factor of $\sim$ 2 larger 
than our adopted rate.  Typical $e$ for leftover planetesimals is probably in this range; thus,
the rate in eq. (\ref{eq: adot-sim}) is somewhat conservative. 

\begin{figure}
\includegraphics[width=6.5in]{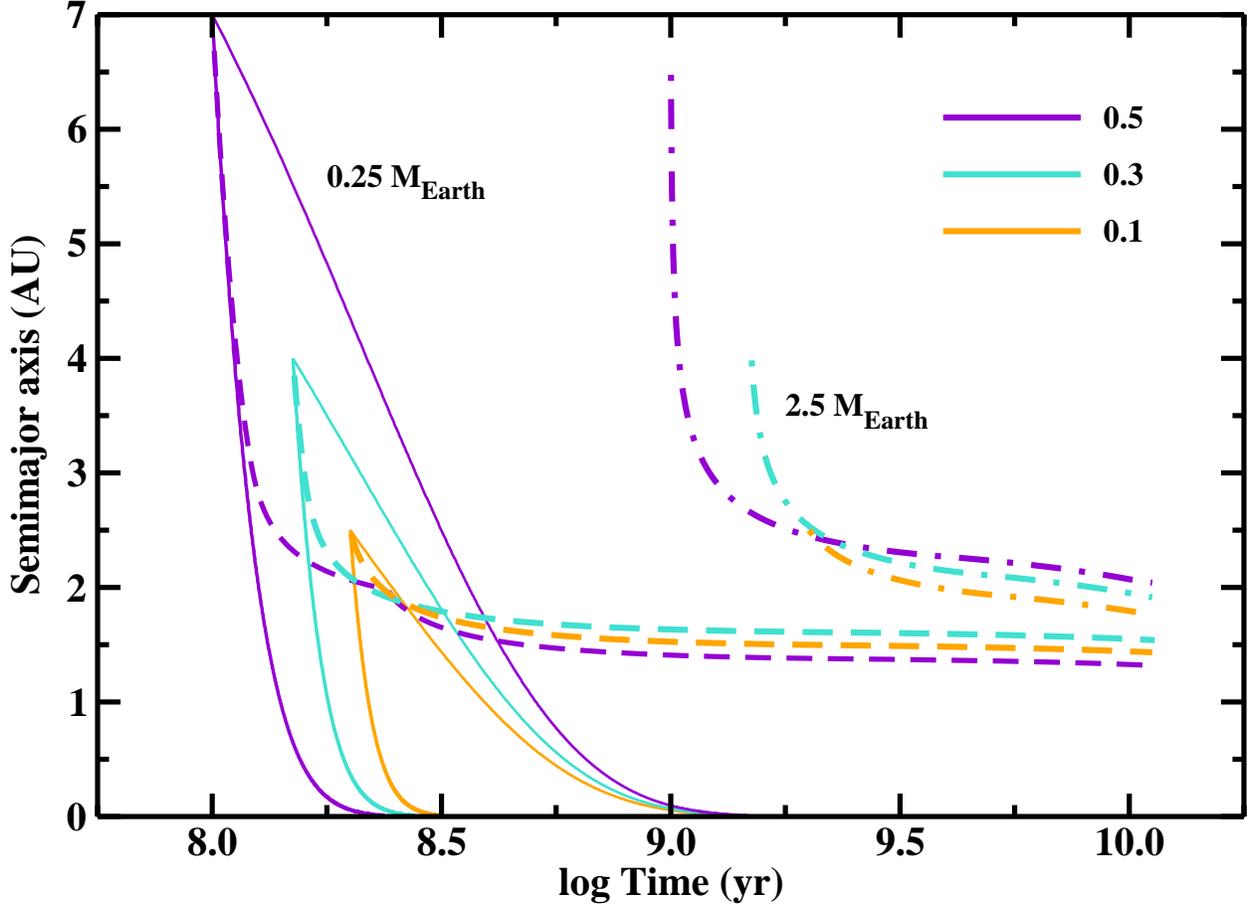}
\vskip 3ex
\caption{%
Semimajor axis evolution for migrating Earth-mass planets orbiting 
0.1~\msun\ (orange curves), 0.3~\msun\ (turquoise curves), or 
0.5~\msun\ (violet curves) stars. Solid curves plot trajectories 
for 0.25~\mearth\ planets drifting through uniformly depleted 
disks with \xm\ = 1 (lower curves) and \xm\ = 0.1 (upper curves).  
Dashed (dot-dashed) curves plot trajectories for 
0.25~\mearth\ (2.5 \mearth) icy planets migrating through disks 
with the more realistic depletion model in eq.~(\ref{eq: xm}). 
In realistic disks, massive planets form later and migrate less 
than lower mass planets.
\label{fig: mig-tracks}
}
\end{figure}

The solid lines in Fig.~\ref{fig: mig-tracks} illustrate the long-term drift of a 
0.25~\mearth\ planet through disks with \xm\ = 0.1 (upper set of three curves) and 
\xm\ = 1.0 (lower set of three curves).  After this planet forms at $t \approx$ 
100--200~Myr, it 
takes from 100~Myr (\xm\ = 1) to 1~Gyr (\xm\ = 0.1) to migrate to within 0.01~AU of 
the host star. In disks with more depletion (\xm\ $\lesssim$ 0.1), the migration
time is longer, $\sim$ 3~Gyr for \xm\ = 0.01 and $\sim$ 10~Gyr for \xm\ = 0.001.
In all cases, however, migration through a disk with constant $\xm \gtrsim 10^{-3}$
results in super-Earth planets very close to the host star.

To explore how migration changes in a disk with a scale factor that depends on
semi-major axis and time, we adopt results from published coagulation calculations
\citep[e.g., \S3;][]{kb2004a,kb2006,kb2008,kb2010}.  We identify the formation 
time for a planet with mass $m$ (e.g., eq. [\ref{eq: t-coll-fin}] and Tables 
\ref{tab: times1}--\ref{tab: times2}) and we infer the scale factor for small 
planetesimals remaining in the disk:
\begin{equation}
\xm\ \approx \left\{
\begin{array}{lll}
1.0 - x_0 (t / t_0 ) & \hspace{5mm} & t < t_0 \\
x_0 ( t / t_0 )^{\epsilon_0} & & t \ge t_0 \\
\end{array}
\right.
\label{eq: xm}
\end{equation}
where $t_0$ is the formation time for the planet and $x_0$ and $\epsilon_0$ are parameters 
that depend on stellar mass and the fragmentation parameters. Typically, 
$x_0 \approx 0.30 - 0.95$ and $\epsilon_0 \approx$ 1. 

Combined with eq.~(\ref{eq: adot-sim}), this expression allows us to track the migration
of a planet through a time-dependent depleted disk of small particles.  After setting 
the initial conditions as a function of stellar mass as outlined above, we then specify the 
stellar mass dependent parameters for eq. (\ref{eq: xm}).  Once these are set, we evolve 
$a$ for a constant mass planet using a simple explicit numerical integration.

Although this model does not include a prescription for the subsequent growth of the planet 
or any dynamical interactions with planets formed at smaller $a$ \citep[e.g.,][]{rogers2011}, 
it captures several important features of migration in a planet forming disk. Because we begin 
the evolution when a planet of mass $m$ forms, the planet begins to migrate at the appropriate 
rate through a disk with the appropriate \xm. By specifying a reasonable \xm\ for the small 
planetesimals which remain in the disk at all $a$, we accurately derive the migration rate 
as the planet migrates closer and closer to the central star. In coagulation models, planets 
form faster closer to the host star.  Thus at any time $t > 0$, the fraction of the initial 
mass remaining in small planetesimals increases outwards 
from the central star. Because the planet migrates through a disk with smaller and smaller 
\xm, it migrates more and more slowly. Our goal is to learn whether the results from our 
numerical calculations of coagulation and migration allow a planet to migrate very close to 
the host star.

In a realistic depleted disk, planets migrate modest distances. When a 0.25~\mearth\ planet
forms at roughly 100--200~Myr, it begins to migrate through a disk where small planetesimals 
contain most of the initial mass. Thus, the planet migrates inward fairly rapidly (Fig.
\ref{fig: mig-tracks}, dashed lines).  As the planet migrates inward, it drifts through
planetesimals with a smaller and smaller fraction of the total mass in solids. Thus, 
migration stalls. For the semi-analytic model outlined above, migration for 
0.25~\mearth\ planets effectively ceases when $a \approx$ 1.5~AU. 

Although more massive planets form later in time, they still migrate close to their parent 
star (Fig. \ref{fig: mig-tracks}, dot-dashed lines).  In disks with identical \xm, a massive 
planet with $m$ = 2.5~\mearth\ migrates roughly 3 times faster than a less massive planet 
with $m$ = 0.25~\mearth\ (eq. [\ref{eq: adot-sim}]).  At the time a more massive planet 
forms, however, there is much less material in small leftover planetesimals.  The much 
smaller mass in planetesimals results in a factor of roughly 10 reduction in migration rate.  
Thus, the 2.5~\mearth\ planet migrates roughly 3 times more slowly than the 0.25~\mearth\ planet.
Migration for this massive planet ends at a larger semimajor axis, $a \approx$ 2~AU. 

The final semimajor axis for migrating icy planets depends on the amount of mass in small
planetesimals remaining in the inner disk. Compared to the nominal model illustrated with
the dashed and dot-dashed lines in Fig.~\ref{fig: mig-tracks}, inner disks with more mass 
in small planetesimals enable more migration. Inner disks with less mass restrict migration 
considerably. For an ensemble of systems with a typical range of mass in small planetesimals 
in the inner disk, we do not generally expect planets to migrate inside of roughly 1~AU.

Despite these general results, some newly-formed icy super-Earths may migrate rapidly to 
semimajor axes of $\sim$ 0.1~AU. In roughly 5\% of our coagulation calculations, collision 
rates `conspire' to produce a super-Earth $\sim$ 20\% earlier than the typical formation 
time. If the disk interior to this super-Earth is normal, then it will migrate 20\% to 40\% 
faster than the typical super-Earth and reach semimajor axes much closer to the central star. 
Modifying our semi-analytic model to treat these outliers yields final semimajor axes of 
0.1--0.2 in 1\% to 3\% of disks with initial \xm\ $\approx$ 1. 

\subsection{Summary of Migration Calculations}

Our discussion leads to several specific conclusions for the migration of icy planets orbiting 
low mass stars.

\begin{itemize}

\item Results for numerical calculations of planets migrating through planetesimal disks generally 
match analytic theory. The numerical rates have the same magnitude as the analytic results and 
scale with disk mass as expected. Scaling with semimajor axis, planet mass, and stellar mass are 
somewhat different from the analytic results. 

\item Earth-mass planets migrate at reasonably large rates, $\sim$ 1~AU~Myr$^{-1}$, 
through a planetesimal disk with a scaled surface density comparable to the minimum mass
solar nebula around a 0.1--0.5 \msun\ star \citep[see also][]{ormel2012}. Even though 
these planets might form quite late in the lifetime of these stars, the rates are 
sufficient to allow the planet to drift close to the star.

\item Combining coagulation and migration calculations into a simple model for the migration of 
an icy Earth-mass planet in an evolving planetesimal disk, we conclude that many Earth-mass 
planets formed at 3--7~AU can migrate to 1--2~AU in 1--3~Gyr. A small fraction of rapidly formed 
super-Earths might migrate to 0.1--0.2~AU. Our model predicts a few icy super-Earths at 0.1--0.2~AU 
for every 100 icy super-Earths at 1--2~AU. 

\item Our simple migration algorithm demonstrates that migration is very sensitive to the mass of 
small planetesimals within the inner disk. Compared to a disk with a power-law surface density and
uniform depletion, planets migrate much smaller distances in a realistic disk where the depletion
increases inversely with semimajor axis.  If planet formation within the inner disk is more efficient
than our nominal model, the inner disk contains little mass in small planetesimals. Icy planets 
fail to migrate. When planet formation is less efficient, however, a more massive inner disk
enables icy planet migration inside 1 AU.

\end{itemize}

These conclusions suggest two formation paths for icy planets close to their parent stars.

\begin{itemize}

\item Early formation channel:  super-Earths form rapidly on time scales of 1--10 Myr. 
These planets migrate close to their parent star while the star contracts on its Hayashi 
track to the main sequence. 
 
\item Late formation channel: super-Earths form slowly on time scales of 1 Gyr. These 
planets migrate when the parent star is on the main sequence.

\end{itemize}

Aside from the identification of two clear formation channels, it is encouraging that our
calculations provide a robust way for planets to migrate from 5--10~AU to 1--2~AU (and 
occasionally to 0.1--0.2~AU). As these planets migrate inward, they probably encounter 
other planets formed closer to the star. Dynamical interactions among these planets may 
lead to the placement of some planets much closer to the parent star 
\citep[e.g.][]{rasio1996,juric2008}. Identifying the outcomes of these interactions requires 
coagulation and migration calculations which cover a larger range of semimajor axes 
(e.g., 0.1--10~AU) than those discussed in \S3--4. Faster parallel computers now enable 
these calculations. Thus, more comprehensive theoretical predictions should be available 
in 1--2 years.

\section{DISCUSSION}
\label{sec: disc}

\subsection{Comparisons with Previous Results}

Current theoretical studies propose a broad range of mechanisms for super-Earth formation 
within a protoplanetary disk.  In the disk instability picture, photoevaporation 
\citep[e.g.,][]{boss2006a} or tidal processes \citep[e.g.,][]{nayak2010} may yield 
super-Earth leftovers from Jupiter-mass protoplanets.  Coagulation models can produce 
rocky super-Earths very close to the host star \citep{mont2009,ogi2009} and icy 
super-Earths somewhat farther from the parent star 
\citep{laugh2004,ida2005,koba2010b,mann2010,rogers2011}.  All of these calculations 
yield super-Earths rapidly, on time scales of 1--10~Myr or less for low mass central 
stars.

In disks composed primarily of large planetesimals, our results identify a much 
slower path to super-Earths. When planetesimals are large, growth is slow. However, 
the collisional cascade is inefficient. Thus, growing protoplanets can accrete nearly 
all of the initial solid mass in the disk. On time scales of roughly 1 Gyr, the 
Mars-mass objects produced during runaway growth evolve into Earth-mass and sometimes 
super-Earth mass objects.

Comparing these results with published calculations is complicated.  Many calculations 
begin with an ensemble of 1000~km or larger objects and follow the evolution with an 
$n$-body calculation 
\citep[e.g.,][]{mont2009,ogi2009,mann2010}. Others consider the history of a single 
Earth-mass protoplanet embedded in a sea of low mass planetesimals and a gaseous disk 
\citep[e.g.,][]{laugh2004,rogers2011}. Although our calculations inform these studies, 
the very different physical approaches preclude robust comparisons. 

Despite the lack of broad comparisons, our results address analytic predictions of 
the velocity equilibrium between viscous stirring and gas drag, the growth time, 
and the final masses of protoplanets. Using similar expressions for velocity evolution 
from gas drag and viscous stirring, the predicted ratio of the equilibrium velocity 
of small particles to the escape velocity of the large particles is roughly 
$v_{esc,l} / v_{eq,s} \propto \Sigma^{-\gamma_1}$, where $\gamma_1$ depends on the 
gas drag regime \citep{raf2004,koba2010b}. Adopting $\gamma_1$ sets the expected 
variation of growth time with surface density (eq. [\ref{eq: t-coll}]),
$t_c \propto \Sigma^{-\gamma_2}$, with $\gamma_2 \approx 2 \gamma_1 + 1$ 
\citep[][Appendix]{kb2008}. The $\gamma_2 \approx 1.1$ inferred from our calculations 
agrees reasonably well with the analytic predictions of \citet{raf2004} and 
\citet{koba2010b}.

To compare our results with analytic predictions for protoplanet masses, we focus 
on derivations consistent with our short gas depletion time.  The final mass 
$M_c$ of a protoplanet which does not accrete fragments produced in a collisional 
cascade \citep[see eq. 21 of][]{koba2010b}
should be similar to the final masses of protoplanets in our simulations. For 100~km
planetesimals at $a$ = 4~AU around a 0.3~\msun\ star, our numerical results for 
an ensemble of calculations with \xm\ = 1 (0.3--1 \mearth) is close to their analytic 
result of 0.7~\mearth. Our prediction for the variation in final mass with the initial 
radius of a planetesimal, $M_f \propto r^{0.15}$, is close to their $M_c \propto r^{0.16}$. 
Given the different approaches, this agreement is remarkable. For more (less) massive 
stars, our derived masses of 0.5--2.1 \mearth\ (0.1--0.26 \mearth) are reasonably close
to the predictions of 0.84 \mearth\ (0.70 \mearth). We suspect the discrepancy at the 
lowest masses is a function of evolution time: evolution times exceeding 10~Gyr allow 
larger final masses in our simulations.

To compare numerical simulations directly, we consider Fig. 8 of \citet{koba2010b},
which shows protoplanet masses at 10~Myr for simulations with $r_0$ = 10~km, 
\xm\ = 1, and $k$ = 3/2 around a solar-mass star. For our simulations, the formation 
time scales as $(\Omega \Sigma)^{-1}$. Thus, we can make an approximate comparison 
using our results for \xm\ = 3 around a 0.5~\msun\ star. At 10~Myr, \citet{koba2010b}
infer masses of roughly 0.04~\mearth\ at 6.5 AU, 0.015~\mearth\ at 9 AU, and
0.003~\mearth\ at 13 AU.  In our ensemble of simulations, we derive
0.03--0.07~\mearth\ at 7 AU, 0.01--0.03~\mearth\ at 9 AU, and
0.002--0.004~\mearth\ at 13 AU. The agreement is satisfactory.

\subsection{Constraints from Observations}

Testing plausible evolutionary paths to super-Earths requires a broad range of 
observations. Here we outline several approaches which could improve our 
understanding of planet formation theory.

\subsubsection{Initial Disk Masses}

Although current data cannot distinguish between the various formation mechanisms for
super-Earths, observations allow robust constraints on the initial masses of solid material 
in protostellar disks.  To make this constraint, we compile the expected frequency 
of Earth-mass or larger planets from our calculations as a function of \xm. For 
calculations with $r_0$ = 30--300~km, the frequency of Earth-mass planets is 
independent of \r0. Thus, we derive an average frequency $\eta$ for this range 
in \r0. To compare with observations, we compile results from microlensing data 
\citep[e.g.,][]{cassan2012} and HARPS radial velocity data \citep{bonfils2013}. 
Both sets of data rely on observations of low mass stars with \mstar\ = 
0.1--0.5~\msun. The microlensing observations are sensitive to 10~\mearth\ and 
larger planets with $a \approx$ 0.5--10~AU. The radial velocity data probe the 
frequency of Earth-mass and larger planets with $a \lesssim$ 0.25~AU.  Together, 
the two samples cover $a \lesssim$ 10~AU and provide a firm lower limit on the 
frequency of Earth-mass and larger planets.

Fig.~\ref{fig: freq} compares the predicted $\eta (x_m)$ with observations.  Colored 
symbols indicate results from our simulations for 0.1~\msun\ (violet), 0.3~\msun\ (blue), 
and 0.5~\msun\ (orange) central stars. Although there is a small increase in planet
frequency with stellar mass \citep[see also][]{kk2008}, there is a marked correlation 
between planet frequency and initial disk mass. Massive disks with \xm\ = 1--3 are much 
more likely to produce Earth-mass and larger planets than disks with \xm\ $\lesssim$ 
0.3. This correlation has a clear reason: low mass disks with \xm\ $\lesssim$ 0.3 do 
not have enough material to form Earth-mass or larger planets.

Observed frequencies of exoplanets (Fig.~\ref{fig: freq}; shaded regions) suggest most 
newly-formed low mass stars are surrounded by massive disks with \xm\ $\gtrsim$ 0.5--1.  
For microlensing and radial velocity measurements, the 1$\sigma$ ranges imply \xm\ $\gtrsim$
1.5 for either set of data. Adopting the 2$\sigma$ range for the microlensing data allows 
all \xm. However, the 3$\sigma$ range for the radial velocity data -- $\eta \gtrsim 0.1$ -- 
rules out disks with \xm\ $\lesssim$ 0.5.  These disk masses are a factor of 5--10 larger 
than the typical masses inferred for disks around pre-main sequence stars with ages of 
1--2~Myr \citep[e.g.,][and references therein]{and2013}.  

The simplest way to reconcile the tension between the disk masses required to explain the 
frequency of super-Earths and those required to explain the mm-wave emission of pre-main 
sequence stars is to postulate significant growth of solids during the earliest phases of 
stellar evolution. In this hypothesis, (i) disks are initially massive and (ii) growth of
micron- to mm-sized grains concentrates most of the initial mass into m-sized or larger
objects in 1--2~Myr. Disk masses derived for pre-main sequence stars then provide an
incomplete measure of the total mass in solids available for planet formation.

Improving these constraints on the initial disk mass requires better estimates for the
frequency of super-Earths and more extensive radio observations to constrain the fraction 
of the total disk mass in m-sized or larger objects. Data from Kepler, MEarth, and ongoing 
microlensing experiments will generate larger populations of super-Earths around M-type
stars \citep[e.g.,][]{berta2012,berta2013}. ALMA observations can improve our 
understanding of the total masses of dusty disks surrounding pre-main sequence stars 
\citep[e.g.,][]{and2013}.

\subsubsection{Planetesimal Formation}

Our calculations also provide a new way to test models of planetesimal formation.  In 
currently popular theories, planetesimals form via coagulation \citep[e.g.,][]{garaud2013}
or some type of dynamical instability \citep[e.g.,][]{johan2012}. Coagulation tends to
concentrate a large fraction of the available solid material into small planetesimals 
with radii of 0.1--10~km. These outcomes serve as starting points for many of our
calculations.  Instabilities within the disk collect 0.1--10~cm pebbles into much 
larger planetesimals with radii of 100~km or larger. In some cases, instabilities may 
produce only a few very large planetesimals which may then accrete pebbles directly 
\citep[e.g.,][see also Bromley \& Kenyon 2011b]{lamb2012}. Our calculations with
large planetesimals address other cases where the instability concentrates most of 
the solid mass in the disk into very large planetesimals.

Although current samples of planets around low mass stars are insufficient, comprehensive 
surveys of low mass stars with a broad range of ages enable additional tests.  When 
super-Earths and gas giants form on short time scales, $\lesssim$ 10--30 Myr, the host 
star is still contracting to the main sequence.  Disks composed only of large planetesimals 
produce super-Earths on much longer time scales, $\gtrsim$ 1~Gyr, after the star reaches the 
main sequence. If both paths to super-Earths operate, low mass stars on the main sequence 
should have a larger fraction of super-Earths than stars approaching the main sequence. 
Defining $f_p$ as the fraction of low mass pre-main sequence stars with super-Earths 
and $f_m$ as the fraction of low mass main sequence stars with super-Earths, 
$f_p / (f_p + f_m)$ ($f_m / (f_p + f_m)$) then represents the fraction of super-Earths 
formed at early (late) times in the evolution of the central star.

\subsubsection{Atmospheric Properties}

Identifying a new formation channel for super-Earths adds another layer of complexity to
predictions for the atmospheric structure of super-Earths. In standard models, icy 
super-Earths form in 1--10 Myr and probably accrete some H-rich gas from the dissipating
circumstellar disk \citep{laugh2004,ida2005,rogers2011}. Rocky super-Earths form later 
when the H-rich disk has almost entirely dissipated \citep{mont2009,ogi2009}. In our new 
formation channel, icy super-Earths form roughly 1~Gyr after the disk has disappeared.
Thus, we expect no accreted atmosphere on icy super-Earths formed at late times from large 
planetesimals.

Despite the lack of accreted atmosphere, icy super-Earths formed at late times may still
have significant atmospheres \citep[e.g.,][]{elkins2008,rogers2010,seager2010,levi2013}. 
As one example, degassing during the late stages of accretion can produce a broad range 
of compositions and total masses for super-Earth atmospheres. Volatiles trapped in ices 
and brought to the surface by convection can add a variety of molecules to the atmosphere.
Predicting the atmospheric structure requires an accurate assessment of the composition 
of icy planetesimals and a detailed understanding of the accretion history.

\section{SUMMARY}
\label{sec: summary}

We describe coagulation calculations of icy planets around 0.1--0.5~\msun\ stars.  Growth 
times for planets as a function of disk mass and semimajor axis match analytic theory, 
$t \propto x_m^{-1.1} a^{k + 3/2}$ where $k$ is the exponent in the power-law relation 
between surface density and semimajor axis. The growth time also depends on the sizes and
intrinsic strengths of the planetesimals that collide and merge into planets. 

These calculations predict a new formation channel for icy super-Earths orbiting low mass 
stars. When icy planetesimals are large ($r_0$ $\gtrsim$ 30--100~km), they grow slowly,
on time scales much longer than typical gas depletion times of a few Myr. With modest 
mass loss due to fragmentation, ensembles of large planetesimals grow into super-Earths
on time scales of $\sim$ 1 Gyr, late in the lifetime of a low mass star.  

Detailed $n$-body calculations suggest Earth-mass planets migrate through remnant planetesimal
disks at rates of $\sim$ 0.01--1~AU~Myr$^{-1}$. Migration rates scale with the disk mass, the
mass of the planet, and the semimajor axis of the planet. The calculated rates agree fairly well 
with those derived from analytic theory.

A simple model combining the results of the coagulation and $n$-body calculations demonstrates
that icy Earth-mass planets can migrate from 5--10~AU to 1--2~AU in $\sim$ 1~Gyr. Lower mass 
planets form earlier and migrate farther than more massive planets. For all Earth-mass planets,
migration from 5--10~AU to 1--2~AU probably leads to dynamical interactions between icy
super-Earths and rocky planets formed closer to the star. These interactions probably place
some planets much closer to their host stars.

\begin{figure}
\includegraphics[width=6.5in]{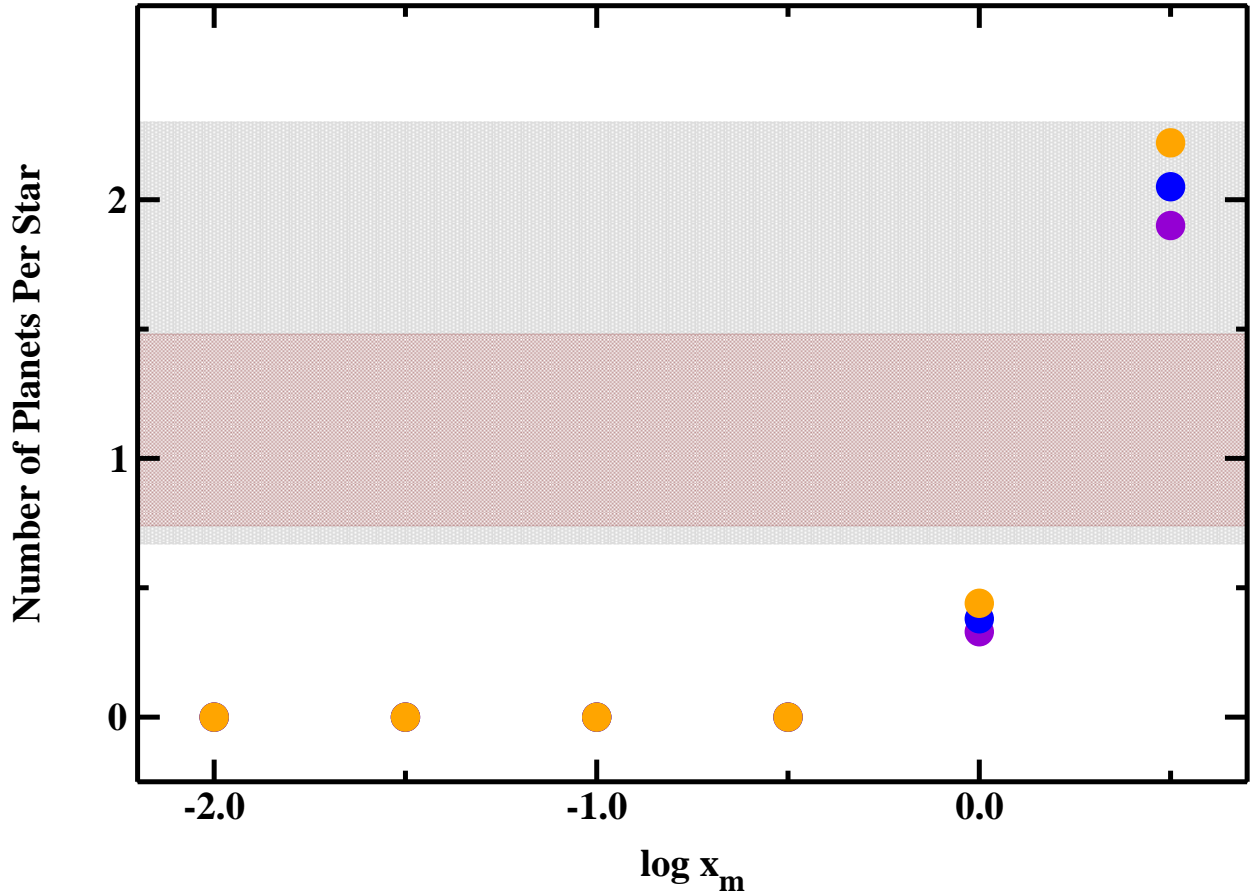}
\vskip 3ex
\caption{%
Comparison of predicted and observed frequencies for Earth-mass and larger exoplanets.
Colored symbols plot predictions for 0.1~\msun\ (violet), 0.3~\msun\ (blue), and 
0.5~\msun\ (orange) stars. The shaded regions show the 1$\sigma$ confidence regions
for exoplanets derived from microlensing data \citep[grey shading;][]{cassan2012} 
and HARPS radial velocity data \citep[brown shading;][]{bonfils2013}. The observations
favor disks with \xm\ $\gtrsim$ 1.
\label{fig: freq}
}
\end{figure}

The high frequency of Earth-mass exoplanets from microlensing and HARPS radial velocity observations 
suggest large initial disk masses for 0.1--0.5~\msun\ stars (Fig.~\ref{fig: freq}). Although the 
exoplanet frequency derived from microlensing data is somewhat larger, the smaller errors in
the frequency derived from the HARPS data provide stronger constraints on initial disk masses.
Both sets of data imply disks with \xm\ $\gtrsim$ 0.5--1.0. 

This result has important implications for the structure of protostellar disks around the 
youngest stars. Observations suggest typical protostellar disks with ages of roughly 1~Myr
have median \xm\ $\lesssim$ 0.1 and dispersions in \xm\ of at least an order of magnitude 
\citep[e.g.,][and references therein]{will2011,and2013}. If the high frequency of super-Earths 
around low mass stars requires disks with initial masses \xm\ $\approx$ 1 and most protostellar 
disks have $\xm \lesssim$ 0.1, then current observations of protostellar disks significantly
underestimate the total mass in solids, implying significant planetesimal growth during the first 
1--2~Myr in the life of a young star \citep[e.g.,][and references therein]{hart1998,furl2009}.

Observations can test other aspects of these calculations.  If super-Earths can form during 
the earliest (1--10~Myr) and latest (1~Gyr) stages in the lfetime of a low mass star, stars 
on the main sequence should have a larger fraction of super-Earths than low mass stars 
contracting on Hayashi tracks towards the main sequence.  Simulations are not yet extensive 
enough to predict a robust semimajor axis range where older stars should have more super-Earths 
than younger stars.  However, identifying an excess for any range of semimajor axes would 
provide strong constraints on planet formation theory.

For low mass stars of any age, better estimates on the mass densities of super-Earth and 
sub-Earth mass planets are important to provide limits on the relative fraction of rocky
and icy/watery planets close to their host stars. Our calculations suggest that some 
planetary systems should contain a mixture of icy and rocky planets. Measuring the 
frequency of these systems will enable important tests of theory.

\vskip 6ex

We acknowledge generous allotments of computer time on the NASA `discover' cluster, 
the SI `hydra' cluster, and the `cosmos' cluster at the Jet Propulsion Laboratory.  
Advice and comments from T. Currie, M. Geller, G. Kennedy, and G. Stewart also 
greatly improved our presentation.  We thank an anonymous referee for a clear and 
thorough review.
Portions of this project were supported by the {\it NASA } {\it Astrophysics Theory} 
and {\it Origins of Solar Systems} programs through grant NNX10AF35G, 
the {\it NASA} {\it TPF Foundation Science Program} through grant NNG06GH25G, 
the {\it Spitzer Guest Observer Program} through grant 20132, and grants from 
the endowment and scholarly studies programs of the Smithsonian Institution.

%\bibliography{master}
%\bibliography{bib2}
\bibliography{ms.bbl}

\clearpage

\begin{deluxetable}{lrrr}
\tablecolumns{4}
\tablewidth{0pc}
\tabletypesize{\footnotesize}
\tablecaption{Initial Disk Masses\tablenotemark{1} (\mearth)}
\tablehead{
  \colhead{} &
  \multicolumn{3}{c}{Stellar Mass in $M_{\odot}$}
\\
  \colhead{$x_m$} &
  \colhead{~~0.1~~} &
  \colhead{~~0.3~~} &
  \colhead{~~0.5~~}
}
\startdata
\cutinhead{$\Sigma_s \propto a^{-1}$}
0.01 & ~~0.07 & ~~0.34 & ~~0.98 \\
0.03 & ~~0.23 & ~~1.12 & ~~3.26 \\
0.10 & ~~0.70 & ~~3.35 & ~~9.79 \\
0.33 & ~~2.33 & ~11.17 & ~32.64 \\
1.00 & ~~7.00 & ~33.51 & ~97.92 \\
\cutinhead{$\Sigma_s \propto a^{-3/2}$}
0.01 & ~~0.01 & ~~0.05 & ~~0.11 \\
0.03 & ~~0.04 & ~~0.17 & ~~0.38 \\
0.10 & ~~0.14 & ~~0.52 & ~~1.15 \\
0.33 & ~~0.46 & ~~1.73 & ~~3.82 \\
1.00 & ~~1.37 & ~~5.20 & ~11.47 \\
3.00 & ~~4.12 & ~15.59 & ~34.41 \\
\enddata
\tablenotetext{1}{Total mass in solid material in the coagulation grid.  
For an adopted gas to dust ratio of 100:1, the total mass of solids and
gas is 100 times larger.}
\label{tab: massgrid}
\end{deluxetable}

\clearpage

\begin{deluxetable}{lccccccccc}
\tablecolumns{10}
\tablewidth{0pc}
\tabletypesize{\footnotesize}
\tablecaption{Maximum planet radii\tablenotemark{1} for calculations with $N_c \propto r^{-3}$}
\tablehead{
  \colhead{} &
  \colhead{} &
  \colhead{} &
  \multicolumn{6}{c}{Disk Mass (\xm)} &
  \colhead{}
\\
  \colhead{\mstar} &
  \colhead{$k$} &
  \colhead{$r_0$ (km)} &
  \colhead{0.01~~} &
  \colhead{0.03~~} &
  \colhead{0.10~~} &
  \colhead{0.33~~} &
  \colhead{1.00~~} &
  \colhead{3.00~~} &
  \colhead{$f_i$} 
}
\startdata
0.1 & 1.0 & 1.0 & 2.91 & 3.02 & 3.15 & 3.28 & 3.40 & \nodata & KB2008 \\
0.3 & 1.0 & 1.0 & 3.02 & 3.15 & 3.27 & 3.43 & 3.58 & \nodata & KB2008 \\
0.5 & 1.0 & 1.0 & 3.10 & 3.25 & 3.39 & 3.51 & 3.65 & \nodata & KB2008 \\
0.5 & 1.5 & 1.0 & 2.97 & 3.08 & 3.19 & 3.35 & 3.47 & 3.61 & KB2008 \\
\\
0.1 & 1.0 & 1.0 & 2.75 & 2.89 & 3.01 & 3.12 & 3.25 & \nodata & LS2009 \\
0.3 & 1.0 & 1.0 & 2.91 & 3.01 & 3.12 & 3.25 & 3.39 & \nodata & LS2009 \\
0.5 & 1.0 & 1.0 & 2.95 & 3.09 & 3.17 & 3.38 & 3.49 & \nodata & LS2009 \\
0.5 & 1.5 & 1.0 & 2.84 & 2.95 & 3.09 & 3.21 & 3.29 & 3.41 & LS2009 \\
\enddata
\tablenotetext{1}{The table lists log \rmax\ where \rmax\ is in km}
\label{tab: rad1}
\end{deluxetable}

\clearpage

\begin{deluxetable}{lccccccccc}
\tablecolumns{10}
\tablewidth{0pc}
\tabletypesize{\scriptsize}
\tablecaption{Maximum planet radii\tablenotemark{1} for calculations with $N_c \propto r^{-0.5}$}
\tablehead{
  \colhead{} &
  \colhead{} &
  \multicolumn{7}{c}{$r_0$ (km)} &
  \colhead{}
\\
  \colhead{\mstar\ (\msun)} &
  \colhead{\xm} &
  \colhead{~~1~~} &
  \colhead{~~3~~} &
  \colhead{~~10~} &
  \colhead{~~30~} &
  \colhead{~100~} &
  \colhead{~300~} &
  \colhead{~1000} &
  \colhead{$f_i$} 
}
\startdata
0.1 & 0.01 & 2.85 & 2.90 & 2.85 & 2.80 & 2.90 & 2.60 & \nodata & LS2009 \\
\nodata & 0.03 & 3.00 & 3.05 & 3.10 & 3.05 & 3.00 & 2.80 & \nodata & LS2009 \\
\nodata & 0.10 & 3.10 & 3.10 & 3.20 & 3.30 & 3.25 & 2.95 & 3.15 & LS2009 \\
\nodata & 0.33 & 3.25 & 3.30 & 3.40 & 3.55 & 3.65 & 3.40 & 3.50 & LS2009 \\
\nodata & 1.00 & 3.40 & 3.45 & 3.60 & 3.80 & 3.80 & 3.90 & 4.05 & LS2009 \\
\nodata & 3.00 & 3.50 & 3.60 & 3.85 & 4.00 & 4.10 & 4.10 & 4.15 & LS2009 \\
\\
0.3 & 0.01 & 2.95 & 3.00 & 3.00 & 2.95 & 2.96 & 3.10 & \nodata & LS2009 \\
\nodata & 0.03 & 3.10 & 3.15 & 3.20 & 3.20 & 3.20 & 3.20 & 3.05 & LS2009 \\
\nodata & 0.10 & 3.25 & 3.30 & 3.35 & 3.40 & 3.45 & 3.40 & 3.10 & LS2009 \\
\nodata & 0.33 & 3.38 & 3.45 & 3.57 & 3.65 & 3.75 & 3.70 & 3.20 & LS2009 \\
\nodata & 1.00 & 3.51 & 3.60 & 3.70 & 3.85 & 4.00 & 4.10 & 4.10 & LS2009 \\
\nodata & 3.00 & 3.64 & 3.75 & 3.86 & 4.23 & 4.13 & 4.20 & 4.20 & LS2009 \\
\\
0.5 & 0.01 & 3.10 & 3.08 & 3.11 & 3.04 & 2.97 & 2.60 & \nodata & LS2009 \\
\nodata & 0.03 & 3.20 & 3.25 & 3.31 & 3.28 & 3.29 & 2.75 & 3.18 & LS2009 \\
\nodata & 0.10 & 3.30 & 3.39 & 3.45 & 3.50 & 3.52 & 3.06 & 3.27 & LS2009 \\
\nodata & 0.33 & 3.46 & 3.55 & 3.62 & 3.75 & 3.80 & 3.40 & 3.40 & LS2009 \\
\nodata & 1.00 & 3.60 & 3.72 & 3.80 & 3.95 & 4.10 & 4.00 & 4.05 & LS2009 \\
\nodata & 3.00 & 3.80 & 3.90 & 4.01 & 4.10 & 4.30 & 4.25 & 4.30 & LS2009 \\
\enddata
\tablenotetext{1}{The table lists log \rmax\ where \rmax\ is in km}
\label{tab: rad2}
\end{deluxetable}
\clearpage

\begin{deluxetable}{lccccccccc}
\tablecolumns{5}
\tablewidth{0pc}
\tabletypesize{\footnotesize}
\tablecaption{Growth Timescales\tablenotemark{1} for calculations with $N_c \propto r^{-3}$}
\tablehead{
  \colhead{\mstar} &
  \colhead{$k$} &
  \colhead{$t_{0,300}$ (Myr)} &
  \colhead{$t_{0,1000}$ (Myr)} &
  \colhead{$f_i$} 
}
\startdata
0.1 & 1.0 & 3.0 & 15 & KB2008 \\
0.3 & 1.0 & 0.8 & ~2 & KB2008 \\
0.5 & 1.0 & 0.3 & ~1 & KB2008 \\
0.5 & 1.5 & 1.0 & ~3 & KB2008 \\
\\
0.1 & 1.0 & 3.0 & 50 & LS2009 \\
0.3 & 1.0 & 0.8 & 10 & LS2009 \\
0.5 & 1.0 & 0.3 & ~2 & LS2009 \\
0.5 & 1.5 & 1.0 & 15 & LS2009 \\
\enddata
\tablenotetext{1}{The table lists coefficients for the relation 
$t_r = t_{0,r} x_m^{-1.1} a_{10}^{n + 3/2}$, where $a_{10}$ is
the semimajor axis in units of 10 AU and $r$ = 300 or 1000. 
The time scales $t_{300}$ and $t_{1000}$ are the time required 
for an ensemble of planetesimals to produce at least one 300~km 
or 1000~km object.}

\label{tab: times1}
\end{deluxetable}

\clearpage

\begin{deluxetable}{lccccccccc}
\tablecolumns{7}
\tablewidth{0pc}
\tabletypesize{\footnotesize}
\tablecaption{Growth Timescales\tablenotemark{1} for calculations with $N_c \propto r^{-0.5}$}
\tablehead{
  \colhead{\mstar} &
  \colhead{$k$} &
  \colhead{\r0} &
  \colhead{$t_{0,300}$ (Myr)} &
  \colhead{$t_{0,1000}$ (Myr)} &
  \colhead{$t_{0,3000}$ (Myr)} &
  \colhead{$f_i$}
}
\startdata
0.1 & 1.5 & 1.0 & 30 & 70 & 30000 & LS2009 \\
0.1 & 1.5 & 3.0 & 60 & 100 & 15000 & LS2009 \\
0.1 & 1.5 & 10.0 & 100 & 250 & 12500 & LS2009 \\
0.1 & 1.5 & 30.0 & 250 & 800 & 11000 & LS2009 \\
0.1 & 1.5 & 100.0 & 150 & 1000 & 11000 & LS2009 \\
0.1 & 1.5 & 300.0 & \nodata & 1500 & 12000 & LS2009 \\
0.1 & 1.5 & 1000.0 & \nodata & \nodata & 11000 & LS2009 \\
\\
0.3 & 1.5 & 1.0 & 7 & 20 & 10000 & LS2009 \\
0.3 & 1.5 & 3.0 & 10 & 25 & 275 & LS2009 \\
0.3 & 1.5 & 10.0 & 20 & 50 & 200 & LS2009 \\
0.3 & 1.5 & 30.0 & 50 & 170 & 300 & LS2009 \\
0.3 & 1.5 & 100.0 & 30 & 200 & 2500 & LS2009 \\
0.3 & 1.5 & 300.0 & \nodata & 250 & 8000 & LS2009 \\
0.3 & 1.5 & 1000.0 & \nodata & \nodata & 12500 & LS2009 \\
\\
0.5 & 1.5 & 1.0 & 3 & 10 & 500 & LS2009 \\
0.5 & 1.5 & 3.0 & 5 & 10 & 30 & LS2009 \\
0.5 & 1.5 & 10.0 & 10 & 15 & 30 & LS2009 \\
0.5 & 1.5 & 30.0 & 15 & 30 & 70 & LS2009 \\
0.5 & 1.5 & 100.0 & 20 & 100 & 400 & LS2009 \\
0.5 & 1.5 & 300.0 & \nodata & 200 & 7000 & LS2009 \\
0.5 & 1.5 & 1000.0 & \nodata & \nodata & 8000 & LS2009 \\
\enddata
\tablenotetext{1}{As defined in Table \ref{tab: times1}. This table includes 
coefficients for $t_{3000}$.}

\label{tab: times2}
\end{deluxetable}

\clearpage

%\centerline{\bf FIGURE CAPTIONS}
%\vskip 4ex
%
\clearpage

\clearpage

\clearpage

\end{document}